\documentclass[final]{siamonline220329}
\usepackage{amsmath, amssymb}
\usepackage{algorithmic}
\usepackage{subcaption}
\usepackage{cleveref}
\usepackage{nicefrac}
\usepackage{xr}
\usepackage{placeins}

\Crefname{ALC@unique}{Line}{Lines}
\title{A generating-function approach to modelling complex contagion on clustered networks with multi-type branching processes\thanks{Submitted to the editors \today. 
\funding{This work was supported by Science Foundation
Ireland [grant numbers 18/CRT/6049 (L.K.), 16/IA/4470 (J.G.), 16/RC/3918 (J.G.), 12/RC/2289 P2 (J.G.)]  with co-funding from the European Regional Development Fund.}}} 
\author{Leah A. Keating\thanks{MACSI, Department of Mathematics and Statistics, University of Limerick, Limerick V94 T9PX, Ireland (\email{leah.keating@ul.ie}, \email{james.gleeson@ul.ie}, \email{david.osullivan@ul.ie}).} \and James P. Gleeson\footnotemark[2] \and David J.P. O'Sullivan\footnotemark[2].} 

\begin{document} 

\maketitle 

\newcommand{\BibTeX}{{\scshapeBib}\TeX\xspace}
\begin{abstract} 
Understanding cascading processes on complex network topologies is paramount for modelling how diseases, information, fake news and other media spread. In this paper, we extend the multi-type branching process method developed in Keating et al., 2022, which relies on homogenous network properties, to a more general class of clustered networks. Using a model of socially-inspired complex contagion we obtain results, not just for the average behaviour of the cascades but for full distributions of the cascade properties. We introduce a new method for the inversion of probability generating functions to recover their underlying probability distributions; this derivation naturally extends to higher dimensions. This inversion technique is used along with the multi-type branching process to obtain univariate and bivariate distributions of cascade properties. Finally, using clique cover methods, we apply the methodology to synthetic and real-world networks and compare the theoretical distribution of cascade sizes with the results of extensive numerical simulations.
\end{abstract} 
\begin{keywords} 
network dynamics, branching processes, complex contagion, probability generating functions
\end{keywords} 
\begin{AMS} 
05C82, 91D30, 60J80, 60J85
\end{AMS}
\section{Introduction}

Recent developments in technology have lead to social network data becoming more available and more analysed than ever before. For example, the Twitter API V2 allows academic researchers to access up to 10 million tweets per month. With this wealth of social network data, we need to develop appropriate tools for its analysis. In this paper, we focus on learning about the interplay between the network structure and the dynamics on the network, such as the spread of behaviour, information or a disease. Social networks are highly clustered \cite{Porter2016DynamicalNetworks, weng2013,watts1998collective,newman_social_2003}, this means that they contain a higher number of triangles than random networks, reflecting the fact that \textit{``a friend of my friend is likely to be a friend of mine''}; however, most models used to model diffusion in online social networks assume that the network is locally tree-like and thus ignoring the clustering \cite{gleeson2021,gleeson2013,Porter2016DynamicalNetworks,melnik2011}. It is important that we develop tools for analysing network dynamics which account for this clustering.
\par
In an online experiment, Centola \cite{Centola2010} observed that repeated exposures to a health behaviour from network neighbours made its adoption more likely, given that the individual had not already adopted, this adoption mechanism is known as a \textit{complex contagion}. Complex contagion dynamics have also been observed in the use of Skype add-ons \cite{karsai2014}, the spread of information and politically-controversial hashtags on Twitter \cite{monsted2017,Romero2011DifferencesTwitter} and the spread of online fads \cite{Sprague2017}. Clustering in the underlying network has been shown to inhibit simple contagion dynamics \cite{shirley2005}, this is because the clustering leads to redundancy in the links for transmission; however, clustering drives the reinforcement mechanism of the complex contagion \cite{Centola2010,OSullivan2015} and therefore larger outbreaks can occur in clustered networks for a complex contagion. Most commonly, complex contagions are modelled using threshold models \cite{watts2002}. In a such models, each node is assigned a threshold and when the number of neighbours who have adopted exceeds the threshold, the node adopts with certainty. In Keating et al.~\cite{2022Keating}, we introduced an alternative model for complex contagion which is an extension of the independent cascade model (ICM) \cite{kempe2003}. We continue to use this model here as it can be easily incorporated into the branching process framework, it allows us to control the effects of social reinforcement through a single parameter, $\alpha\in\left[0,1\right]$, and if we set $\alpha=0$ we get the ICM of \cite{kempe2003} which is a simple contagion; i.e, we can study both simple and complex contagions through this model.
\par
Branching processes are discrete-time stochastic processes that have been used to model diffusion cascades \cite{Gleeson2014,gleeson2016effects,gleeson2014competition,gleeson2017avalanches,Aragon2017GenerativeChallenges,McSweeney2020}, the spread of infectious diseases \cite{vazquez06,vazquez21} and population dynamics in ecology \cite{Caswell2018MatrixModels}. Simple branching processes have proven a fruitful means of capturing important network properties, Gleeson et al.~\cite{gleeson2021} analytically derived cascade properties from data including the cascade size distribution, cascade lifetime distribution, expected average tree depth (EATD) and structural virality \cite{goel2016structural}. The authors focused on the case where the network is assumed to be locally tree-like; i.e., unclustered, and considered simple-contagion dynamics only. The branching process theory for modelling cascades on networks assumes an infinitely large network; however, as has been shown in numerous previous papers \cite{gleeson2021,2022Keating,obrien2020} and in this paper, branching processes can give very accurate results even when the network size is clearly finite. In previous work \cite{2022Keating}, we developed a method for modelling complex contagion on clustered networks using multi-type branching processes (MTBPs). We concentrated on average measures of the cascades and analytically calculated the cascade condition and the expected cascade size for a very specific class of networks. In this paper, for clustered networks, we analytically derive results including the complete distribution of cascade sizes, the distribution of cascade lifetimes and the EATD. We also extend the MTBP theory to more general network distributions, making the method more applicable to real-world networks. In particular, we use a family of network distributions proposed independently by Newman \cite{Newman2009} and Miller \cite{Miller2009} in 2009; described by the distribution of triangle and single-link membership of the nodes, we refer to these networks as Newman-Miller networks. By incorporating clustering into the model through the Newman-Miller networks, we can analytically study both simple- and complex-contagion spreading processes.
\par
Within the branching process framework, we use probability generating functions (pgfs) to derive distributions of the quantities of interest. When we have a pgf for a quantity such as cascade size, we are interested in recovering its full probability distribution from the pgf. The most common method of numerically finding the probability distribution in the network dynamics literature is that of Cavers \cite{cavers1978,obrien2020,gleeson2014competition}, this involves using the z-transform inversion which has the form of a contour integral and approximating the integral using the trapezoidal rule. For bivariate pgfs Brummitt et al.~\cite{brummit2012} use three different methods of finding the probability distribution in a two-type branching process, but these methods use computer algebra systems and are restricted in the number of terms that they can retrieve. While Cavers' derivation has been extended to bivariate pgfs \cite{Antal2010}, it is not clear how it extends to higher dimensions. In this paper, we introduce an alternative derivation of the pgf inversion method. Our approach differs to that of Cavers \cite{cavers1978} and Antal \cite{Antal2010}; while they focus on the z-transform inversion, we look at the equivalence between the pgf evaluated at specific points and the discrete Fourier transform of the probability distribution. Our method has the same computational implementation as Cavers' method in one dimension but has the added advantage of naturally extending to the inversion of pgfs of any dimension. One of the contributions of this paper is to describe and show how to implement the inversion of univariate and bivariate pgfs, the same method may be straightforwardly extended to higher-dimensional distributions.
\par
We are interested in understanding how the structure of a network and the dynamics interact with each other. It is important that we can apply these methods to real-world networks. Burgio et al.~\cite{burgio21} introduced a clique cover method which allows us to approximate the (maximal) clique-membership distribution of a network while constraining that the cliques are edge-disjoint --- cliques do not share edges --- called the edge-disjoint edge clique cover (EECC). In \cref{sec:app_to_data} we apply the EECC to synthetic and real-world networks and find the cascade size distribution according to the branching process theory, for comparison with Monte-Carlo simulations. Applying the MTBP to these empirical networks allows us to see where the theory gives accurate results and to identify areas for future work on the branching process methodology.
\par
The rest of this paper is structured as follows; in \cref{sec:adoption_dynamics} we describe the complex contagion adoption dynamics, in \cref{sec:newman_miller} we introduce the class of networks that we use in our calculations, in \cref{sec:cascade_properties} we show how the MTBP framework can be leveraged to derive distributions of cascade properties. Our examples in \cref{sec:cascade_properties} include the distributions of cascade size and cascade lifetimes and the joint distribution of cascade size and cumulative depth. In \cref{section:1D_inversion} we derive a method for the inversion of pgfs to recover the probability distribution, in \cref{sec:app_to_data} we discuss and show results for applying this method to real-world networks and in \cref{sec:discussion} we discuss the capabilities and limitations of our approach.

\section{Complex Contagion Adoption Dynamics\label{sec:adoption_dynamics}}

When we study a diffusion process on a network, both the spreading mechanism and the network structure are required to accurately model the process. To model complex contagion using the MTBP methodology, we use the discrete-time model of \cite{2022Keating}, which is an extension of the ICM \cite{kempe2003}. Nodes in the network are assumed to be in one of three states; active, inactive or removed. The spreading process is initiated with a small number of active seed nodes that are randomly selected --- in examples here we choose to select just one seed node, all other nodes are initially inactive. At each time step, active nodes have one chance to activate each of their inactive neighbours, activation occurs with probability $p_{k}$, where $k$ is the number of times that that inactive neighbour has been exposed. If the activation is successful, the inactive neighbour becomes active in the next time step, active nodes always become removed in the following time step and once a node enters the removed state, it never leaves that state. The dynamics are governed by two parameters; $p_{1}$ and $\alpha$, where $p_{1}\in \left[0,1\right]$ is the probability of adoption after a single exposure and $\alpha\in\left[0,1\right]$ is a measure of the strength of social reinforcement. Higher values of $\alpha$ represent higher levels of social reinforcement and in the limiting case of $\alpha=0$ we recover exactly the (simple-contagion) independent cascade model. The probability that a node will become active after its $k^{th}$ exposure, given that it is inactive, is given by the following equations,
\begin{equation}
    p_{k} = 1-q_{k}
\end{equation}
and
\begin{equation}
    q_{k} = q_{1}(1-\alpha)^{k-1},
\end{equation}
where $q_{k}$ is the probability that a node does not adopt directly after the $k^{th}$ exposure, given that it has not already adopted. This model was studied in \cite{2022Keating} for a small number of networks with homogeneous Newman-Miller distributions. Some of the results included the cascade condition and the expected cascade size.

\section{Newman-Miller Distributions\label{sec:newman_miller}}

Along with the complex contagion adoption dynamics that we described in the previous section, to fully describe the MTBP we need a way of describing the network so that it can be incorporated into the MTBP. We model diffusion on networks where the clique membership of the nodes follows a joint probability distribution $\pi_{st}$, which is defined as the probability that a node, chosen at random, is in $s$ single links and $t$ triangles (is in $s$ 2-cliques and $t$ 3-cliques). We refer to $s$ as the \textit{link degree} and $t$ as the \textit{triangle degree}. This family of networks was independently proposed by Newman \cite{Newman2009} and Miller \cite{Miller2009} in 2009 and we have elected to use these network models because they are commonly used to analytically study spreading processes on clustered networks \cite{Mann2022,McSweeney2020}. We use pgfs to describe the joint triangle-link (or Newman-Miller) degree distribution of the network. The pgf is defined as
\begin{equation}
    \tilde{f}\left(x,y\right) = \sum_{s,t}\pi_{st}x^{s}y^{t}.\label{eq:netpgf}
\end{equation}
Here, we use the example of the doubly-Poisson distribution \cite{Newman2009} which has probability mass function (pmf)
\begin{equation}
    \pi_{st} = e^{-\mu}\frac{\mu^{s}}{s!} e^{-\nu}\frac{\nu^{t}}{t!},
\end{equation}
where $\mu$ is the expected number of single links per node and $\nu$ is the expected number of triangles per node. We can express this distribution in the form of a pgf using Eq.~\cref{eq:netpgf},
\begin{equation}
    \tilde{f}\left(x,y\right) = \sum_{s,t=0}^{\infty}e^{-\mu}\frac{\mu^{s}}{s!} e^{-\nu}\frac{\nu^{t}}{t!}x^{s}y^{t} = e^{\mu(x-1)+\nu(y-1)}.
\end{equation}
When we explore the dynamics of the system, we often need to know the excess degree distributions from traversing a link or a triangle, to allow us to calculate the cascade size distribution, for example. If we traverse a single link then the joint excess degree is the number of triangles and single links that the node at the other end is a part of, not including the single link traversed itself. If we traverse a triangle, the joint excess degree is the number of single links and triangles that the node at the other end is part of, not including the triangle traversed. As described by Newman \cite{Newman2009}, we can find the joint excess triangle-link distributions from traversing a single link $f_{q}$ and from traversing a triangle $f_{r}$ as follows;
\begin{equation}
    f_{q} = \frac{\sum_{st}s\pi_{st}x^{s-1}y^{t}}{\sum_{st}s\pi_{st}}= \frac{\nicefrac{\partial}{\partial x} \tilde{f}(x,y) }{\nicefrac{\partial}{\partial x} \tilde{f}(x,y)|_{x=y=1}}\label{eq:fq}
\end{equation}
and
\begin{equation}
    f_{r} = \frac{\sum_{st}t\pi_{st}x^{s}y^{t-1}}{\sum_{st}t\pi_{st}}= \frac{\nicefrac{\partial}{\partial y} \tilde{f}(x,y) }{\nicefrac{\partial}{\partial y} \tilde{f}(x,y)|_{x=y=1}}.\label{eq:fr}
\end{equation}
In the following sections, we use these network-structure distributions in our calculations of the distributions and values of quantities of interest.

\section{Derivations of cascade properties\label{sec:cascade_properties}}

In this section, using pgfs, we theoretically derive distributions of cascade size, cascade lifetimes and the joint distribution of cascade size and cumulative depth. We also use the moments of the joint distribution to calculate the Pearson's correlation coefficient between cascade size and cumulative depth and the expected average tree depth (EATD).

\subsection{Cascade Size Distribution\label{sec:size_dist}}

\begin{figure}[h]
    \centering
\begin{subfigure}{\textwidth}
    \centering
    \includegraphics[width = \textwidth]{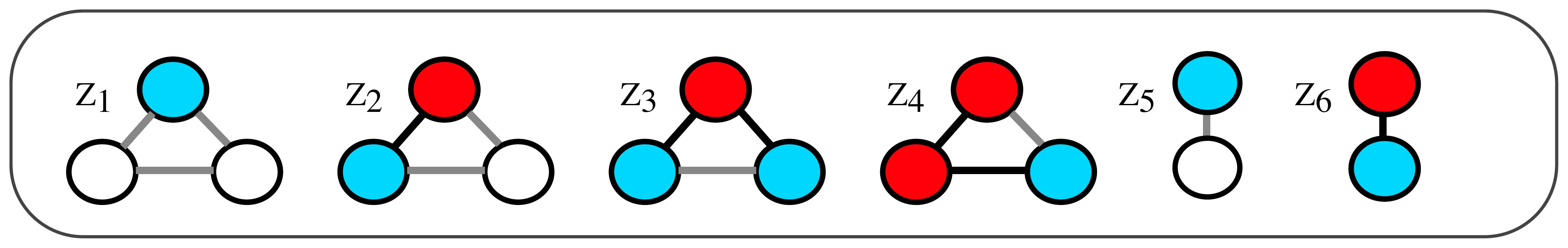}
    \caption{}
\end{subfigure}
\centering
\begin{subfigure}{\textwidth}
    \centering
    \includegraphics[width = \textwidth]{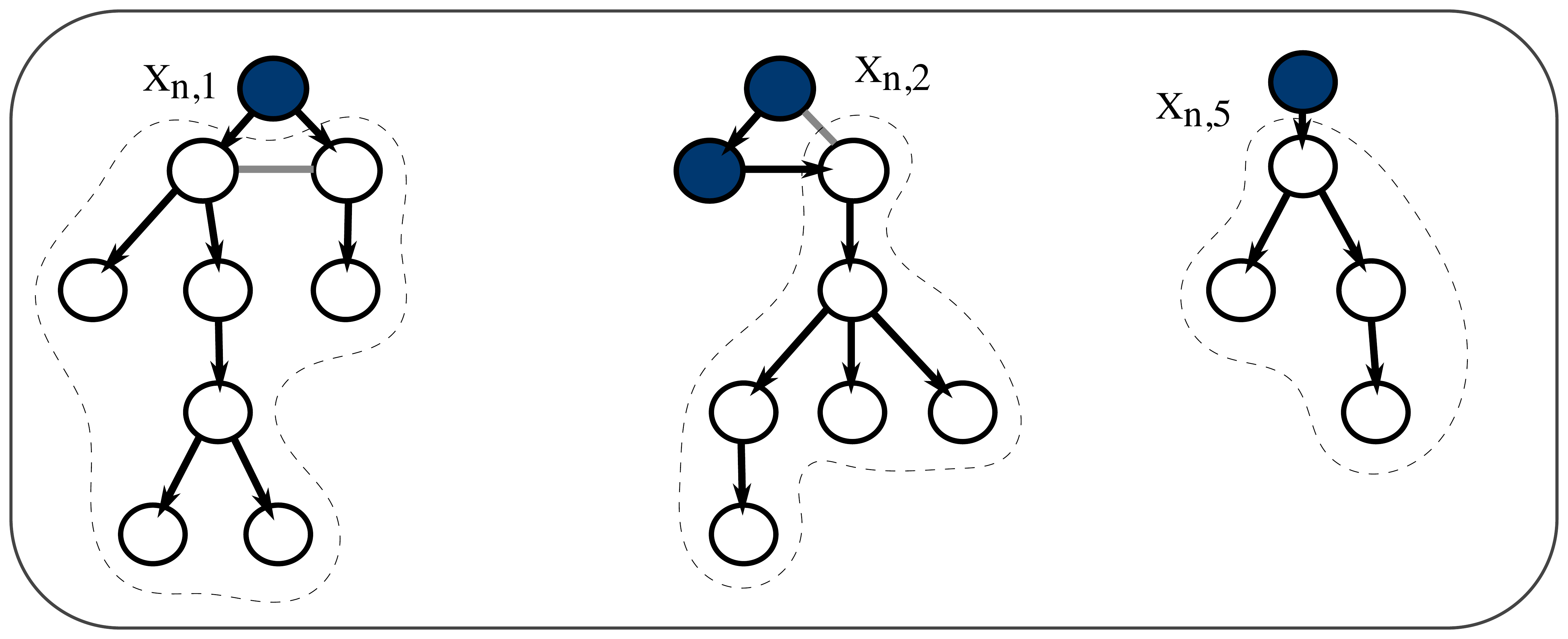}
    \caption{}
\end{subfigure}
    \caption{(a) The 6 possible clique motifs in a Newman-Miller network, blue nodes are active, white nodes are inactive and red nodes are removed. (b) A visualisation of the node counting in subtrees for cascade size, we do not count active or removed nodes in the seed motif (dark blue) and only count all other nodes in the subtree (in white). In this figure we assume that the number of generations that we allow the subtree to grow for $n\to\infty$. We let $n\to\infty$ because, for a sub-critical branching process, this assures that the sub-trees will be fully grown; i.e., there will be no further offspring in future generations.\label{fig:motifs_subtrees}}
\end{figure}

Here, we analytically derive the cascade-size distribution for complex contagion dynamics on a network constructed using the Newman-Miller approach described in \cref{sec:newman_miller}. Previously, we have calculated the average cascade behaviour using MTBPs \cite{2022Keating}. The \textit{types} in the MTBP are \textit{clique motifs}, as shown in \cref{fig:motifs_subtrees} (a), a clique motif is a clique with a specific number of active, inactive and removed nodes. We are only concerned with those motifs with at least one active node. In a Newman-Miller network there are 6 possible clique motifs, we denote by $z_{i}$ a type-$i$ motif. In the paper \cite{2022Keating}, we showed how to calculate the expected cascade size; however, often it is the distribution of cascade sizes that is of interest, not just the expected value. Other works analytically derive this distribution for locally tree-like networks \cite{gleeson2021,gleeson2016effects,gleeson2014competition} but not for clustered networks. Here, we derive the distribution of cascade sizes for the complex contagion dynamics that we described in \cref{sec:adoption_dynamics} on (clustered) Newman-Miller networks.
\par
We consider the size of a full cascade after $n$ generations, $\tilde{X}_{n}$, in terms of its subtree sizes, where the size of a type-$i$ subtree after $n$ generations, $X_{n,i}$, is the number of nodes in the tree seeded by a type-$i$ motif that were once or are currently active or removed, excluding those active or removed nodes that were in the seed motif; this is illustrated in \cref{fig:motifs_subtrees} (b). Not including these active or removed nodes in the seed motif avoids counting a node in the full cascade more than once. 
\par
\begin{figure}[h]
    \centering
         \begin{subfigure}{0.49\textwidth}
         \centering
         \includegraphics[width=\textwidth]{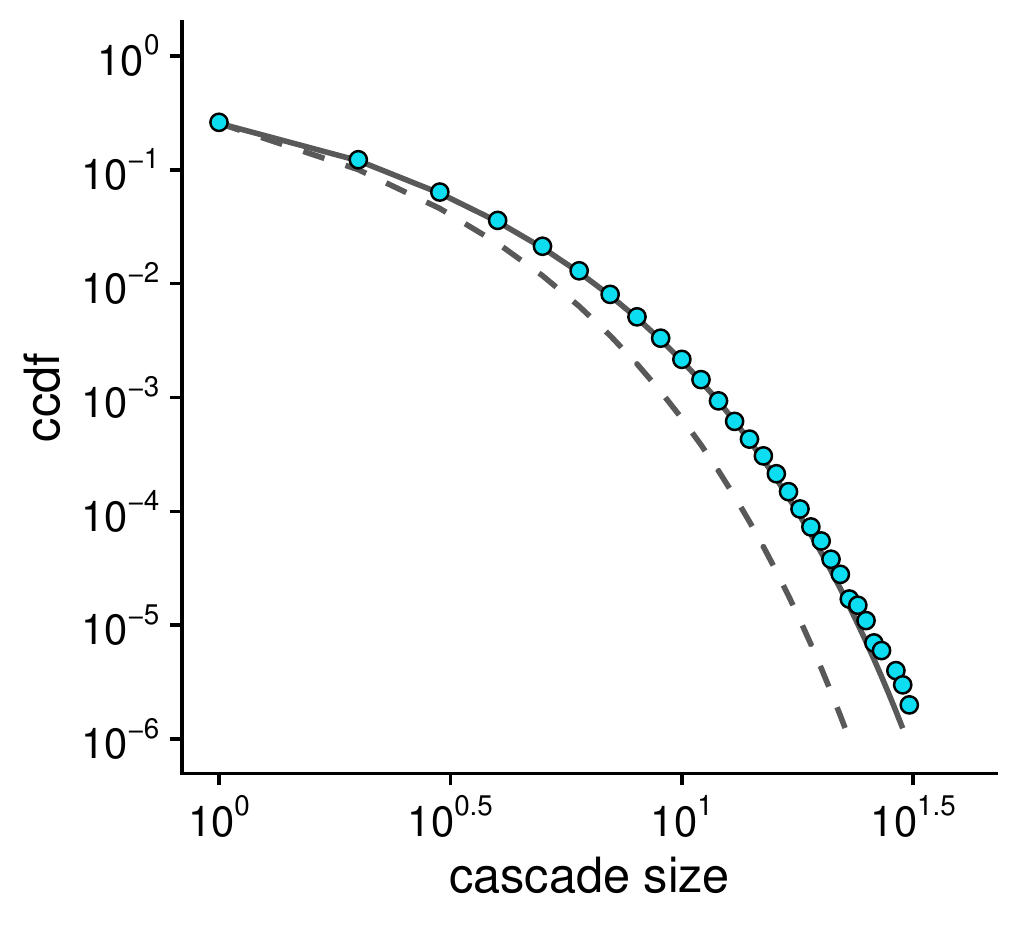}
         \caption{}
     \end{subfigure}
     \hfill
    \centering
             \begin{subfigure}{0.49\textwidth}
         \centering
         \includegraphics[width=\textwidth]{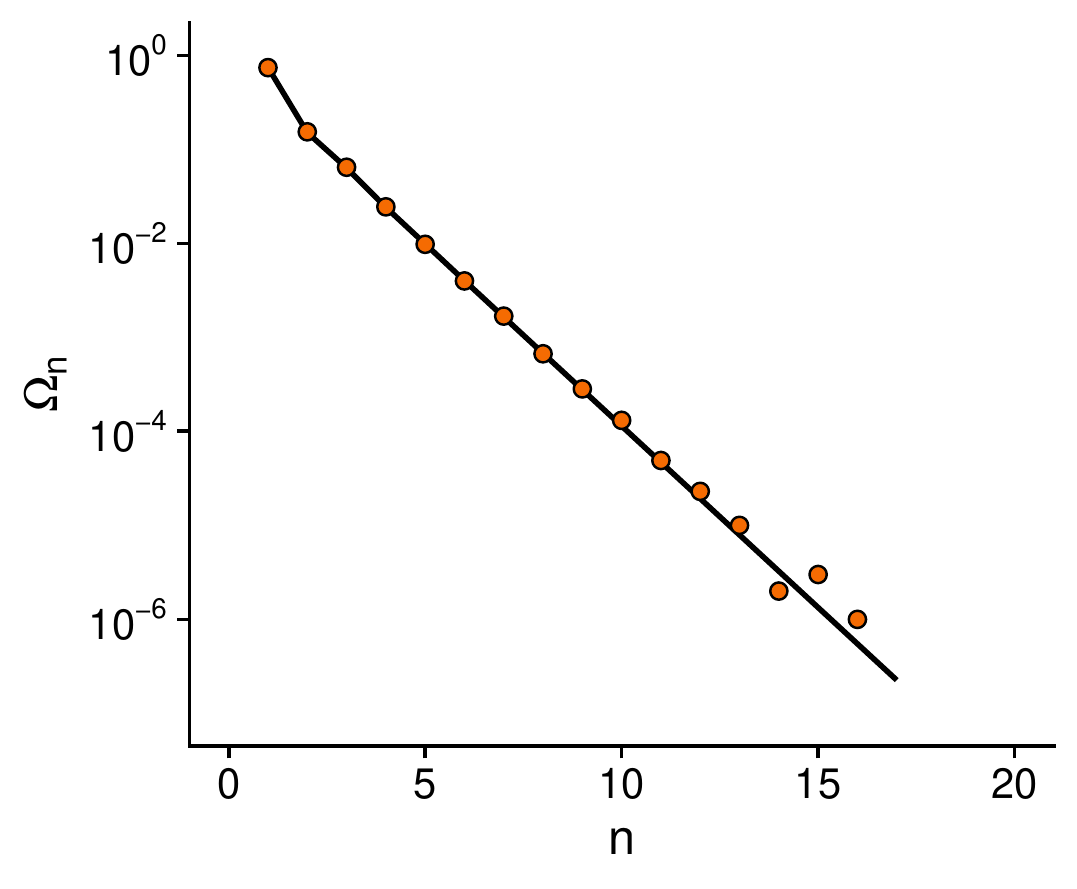}
         \caption{}
     \end{subfigure}
    \caption{(a) The theoretical complimentary cumulative distribution function (CCDF) of cascade size from the MTBP (solid line) and for a simple branching process (dashed line) and 1 million simulations on a network of size 10,000 (blue points). The CCDF for a random variable $Y$ is defined as $P\left(Y>y\right)$, the probability that the random variable $Y$ is greater than a specific $y$. (b) The probability distribution of cascade lifetimes where $\Omega_{n}$ is the probability that the cascade survives to generation $n$. The solid line represents the theoretical distribution and the orange points are the values from simulations. The results shown are for doubly-Poisson Newman-Miller networks with $\mu = 2$ and $\nu = 2$ (average degree 6) and for parameters $p_{1} = 0.05$ and $\alpha = 0.2$.}
    \label{fig:cascade_size_dist}
\end{figure}

To find the cascade-size distribution, we start by finding the pgf for cascade size. By definition, the pgf for cascade size for a full tree that has grown from its seed for $n$ generations, $\tilde{K}_{n}(z)$, is
\begin{equation}
    \tilde{K}_{n}(z) = \sum_{l=0}^{\infty}P(\Tilde{X}_{n}=l)z^{l}= \mathbb{E}\left[z^{\tilde{X}_{n}}\right],
\end{equation}
where $P(\Tilde{X}_{n}=l)$ is the probability that the size of the full cascade $\tilde{X}_{n}$ is $l$. The full cascade size, $\tilde{X}_{n}$, can be expressed in terms of its subtree sizes,
\begin{equation}
    \tilde{X}_{n} = 1 + \sum_{i = 1}^{s}X_{n,5}^{(i)} + \sum_{j = 1}^{t}X_{n,1}^{(j)},
\end{equation}
where $s$ is the number of subtrees seeded with a type-5 motif and $t$ is the number of subtrees seeded with a type-1 motif --- the seed node is in $t$ triangles and has $s$ single links (see \cref{fig:motifs_subtrees} (a) for the motif types). The pgf for the full cascade size after $n$ generations, $\tilde{K}_{n}(z)$, is given by
\begin{align}
        \tilde{K}_{n}(z) &=\mathbb{E}\left[z^{\tilde{X}_{n}}\right]\label{eq:size_pgf_full}\\
        &=\sum_{s,t}\pi_{st}\mathbb{E}\left[z^{1 + \sum_{i = 1}^{s}X_{n,5}^{(i)} + \sum_{j = 1}^{t}X_{n,1}^{(j)}}\middle| \text{ seed node has }s\text{ links and }t\text{ triangles }\right] \nonumber\\
        &= z\sum_{s,t}\pi_{st}\mathbb{E}\left[z^{X_{n,5}}\right]^{s}\mathbb{E}\left[z^{X_{n,1}}\right]^{t}\nonumber\\
        &=z\tilde{f}\left(K_{n,5}(z),K_{n,1}(z)\right).\nonumber
\end{align}
To find $\tilde{K}_{n}(z)$, we need to know the pgfs for the subtree sizes $K_{n,1}(z)$, $K_{n,2}(z)$ and $K_{n,5}(z)$. The pgf for a type-1 subtree, after $n$ generations is given by,
\begin{align}
    K_{n,1}(z) &= \mathbb{E}\left[z^{X_{n,1}}\right]\label{eq:cascade_subtree1}\\
    &=\sum_{l = 0}^{2}\binom{2}{l}p_{1}^{l}(1-p_{1})^{2-l}\mathbb{E}\left[z^{X_{n,1}}\middle| ~l\text{ active nodes in generation }1\right]\nonumber\\
    &=(1-p_{1})^{2}z^{0}\nonumber\\
    &+2p_{1}(1-p_{1})\mathbb{E}\left[z^{1+X_{n-1,2}+\sum_{i}X_{n-1,1}^{(i)}+\sum_{j}X_{n-1,5}^{(j)}}\right]\nonumber\\
    &+p_{1}^{2}\mathbb{E}\left[z^{2+\sum_{i}X_{n-1,1}^{(i)}+\sum_{j}X_{n-1,5}^{(j)}+\sum_{k}X_{n-1,1}^{(k)}+\sum_{l}X_{n-1,5}^{(l)}}\right]\nonumber,
\end{align}
where $X_{n,1}$ is a subtree seeded by a type-1 motif. The three terms in Eq.~\cref{eq:cascade_subtree1} come from the fact that a type-1 motif, in the next generation, either gives rise to no new active nodes, one new active node or two new active nodes respectively. These new active nodes must be counted in the cascade size along with the generation $n-1$ subtrees that they seed. For example, if in generation 1 a type-1 motif has 1 offspring; i.e, produces a type-2 motif, the subtree size can be written in the form, $1+X_{n-1,2}+\sum_{i}X_{n-1,1}^{(i)}+\sum_{j}X_{n-1,5}^{(j)}$, where the 1 counts the new active node, $X_{n-1,2}$ is for the size of the type-2 subtree produced and the $\sum_{i}X_{n-1,1}^{(i)}+\sum_{j}X_{n-1,5}^{(j)}$ is for the generation $n-1$ type-1 and type-5 subtrees produced, where the number of type-1 subtrees is the excess triangle degree from traversing a triangle and the number of type-5 subtrees is the excess link degree from traversing a triangle. We can rewrite Eq.~\cref{eq:cascade_subtree1} as
\begin{align}
    K_{n,1}(z) &= (1-p_{1})^{2}\\
    &+2p_{1}(1-p_{1})z\mathbb{E}\left[z^{X_{n-1,2}}\right]\sum_{s,t}r_{st}\mathbb{E}\left[z^{X_{n-1,5}}\right]^{s}\left[z^{X_{n-1,1}}\right]^{t}\nonumber\\
    &+p_{1}^{2}z^{2}\left(\sum_{s,t}r_{st}\mathbb{E}\left[z^{X_{n-1,5}}\right]^{s}\left[z^{X_{n-1,1}}\right]^{t}\right)^{2},\nonumber
\end{align}
which becomes
\begin{align}
    K_{n,1}(z) &=(1-p_{1})^{2} + 2p_{1}(1-p_{1})zK_{n-1,2}(z)f_{r}\left(K_{n-1,5}(z),K_{n-1,1}(z)\right)\nonumber\\
    &+p_{1}^{2}z^{2}\left(f_{r}\left(K_{n-1,5}(z),K_{n-1,1}(z)\right)\right)^{2}\nonumber
\end{align}
when we notice that these expectations and sums are equivalent to the pgfs for the offspring distribution from traversing a triangle, $f_{r}(x,y)$, and the pgfs for subtree size after $n-1$ generations. Similarly, by using the offspring distributions from a type-2 and a type-5 motif we can get the subtree size pgfs for type-2 and type-5 subtrees:
\begin{align}
    K_{n,2}(z) &= \mathbb{E}\left[z^{X_{n,2}}\right]\\
    &= 1-p_{2} + p_{2}zf_{r}\left(K_{n-1,5}(z),K_{n-1,1}(z)\right),\nonumber
\end{align}
and
\begin{align}
    K_{n,5}(z) &= \mathbb{E}\left[z^{X_{n,5}}\right]\label{eq:size_pgf_6}\\
    &= 1-p_{1} + p_{1}zf_{q}\left(K_{n-1,5}(z),K_{n-1,1}(z)\right).\nonumber
\end{align}
Initially (in generation $n=0$) the subtree size is zero; i.e., $X_{0,1}=X_{0,2}=X_{0,5}=0$, giving us initial conditions for their respective pgfs of $\mathbb{E}\left[z^{X_{0,i}}\right]=\mathbb{E}\left[z^{0}\right]=1$. We iterate through Eqns.~\cref{eq:size_pgf_full} to \cref{eq:size_pgf_6} with initial conditions; $K_{0,1}(z)=K_{0,2}(z)=K_{0,5}(z)=1$, to find $\tilde{K}_{n}(z)$. With this, we have a system of equations that describe the pgf for cascade size; in \cref{section:1D_inversion} we show how to recover the probability distribution from the pgf, allowing us to plot the cascade-size distribution for a given Newman-Miller network with given values for $p_{1}$ and $\alpha$ in the adoption dynamics. In \cref{fig:cascade_size_dist} we show an example of the cascade size distribution derived from the pgf compared to simulations on a network with a doubly-Poisson distribution. In \cite{2022Keating} we calculated the expected cascade size for sub-critical cascades, here we calculate the distribution of cascade sizes for sub-critical cascades in the limit as the number of generations $n\to\infty$. The cascades are \textit{sub-critical} if an infinitesimally small seed fraction of active nodes will not generate cascades of non-finite mean size, which is a non-vanishing fraction of the network as the network size tends to the large network limit. To find the pgf in the limit as $n\to\infty$, we set a tolerance of $10^{-5}$ such that we stop iterating once the pgf evaluated at generation $n$ is within $10^{-5}$ of the pgf evaluated at generation $n-1$.

\subsection{Distribution of cascade lifetimes\label{sec:cascade_lifetimes}}

Here, we show how to find the distribution of cascade lifetimes. The lifetime of a cascade is the number of generations it lives for before dying out; i.e., the lifetime is the maximum of the generations of all nodes present in the cascade. If a cascade has particles --- by particle we mean an active node --- in generation $n-1$ but no particles in generation $n$, the lifetime of the cascade is $n$. To calculate the probability $\Omega_{n}$ that the lifetime of a cascade is exactly $n$ generations, we therefore find the probability that the number of particles in generation $n$ is zero and subtract the probability that there are zero particles in generation $n-1$. Let the number of particles in generation $n$ be given by $\Tilde{Z}_{n}$, and define the corresponding pgf to be
\begin{equation}
    \tilde{F}_{n}(x) = \sum_{l=0}^{\infty}P(\Tilde{Z}_{n}=l)x^{l}= \mathbb{E}\left[x^{\tilde{Z}_{n}}\right].
\end{equation}
We write the pgf in terms the number of particles in generation $n$ of its subtrees
\begin{align}
    \tilde{F}_{n}(x)&=\sum_{s,t}\pi_{st}\mathbb{E}\left[x^{\sum_{i = 1}^{s}Z_{n,5}^{(i)} + \sum_{j = 1}^{t}Z_{n,1}^{(j)}}\middle| \text{ seed node has }s\text{ links and }t\text{ triangles }\right]\\
    &=\sum_{s,t}\pi_{st}\mathbb{E}\left[x^{Z_{n,5}}\right]^{s}\mathbb{E}\left[x^{Z_{n,1}}\right]^{t}\nonumber\\
    &=\tilde{f}\left(F_{n,5}(x),F_{n,1}(x)\right),\nonumber
\end{align}
for $n\geq1$, where $F_{n,5}(x)$ and $F_{n,1}(x)$ are the pgfs for the number of particles in generation $n$ for type-5 and type-1 subtrees respectively. For completeness, if $n=0$ then $\tilde{F}_{0}(x) = \mathbb{E}\left[x^{\tilde{Z}_{0}}\right]= \mathbb{E}\left[x^{1}\right]=x$. To fully compute $F_{n}(x)$, we need $F_{n,1}(x)$, $F_{n,2}(x)$ and $F_{n,5}(x)$, which we calculate in a similar way to the pgfs for subtree size in \cref{sec:size_dist},
\begin{align}
    F_{n,1}(x) = &(1-p_{1})^{2}+2p_{1}(1-p_{1})F_{n-1,2}(x)f_{r}\left(F_{n-1,5}(x),F_{n-1,1}(x)\right)\\
    & p_{1}^{2}f_{r}^{2}\left(F_{n-1,5}(x),F_{n-1,1}(x)\right)\nonumber,
\end{align}
\begin{equation}
    F_{n,2}(x) = 1-p_{2} + p_{2}f_{r}\left(F_{n-1,5}(x),F_{n-1,1}(x)\right)
\end{equation}
and
\begin{equation}
    F_{n,5}(x) = 1-p_{1} + p_{1}f_{q}\left(F_{n-1,5}(x),F_{n-1,1}(x)\right),
\end{equation}
with the initial conditions:
\begin{equation}
    F_{1,1}(x)=(1-p_{1})^{2}+2p_{1}(1-p_{1})x+p_{1}^{2}x^{2},\label{eq:IC_lifetime_1}
\end{equation}
\begin{equation}
    F_{1,2}(x)=1-p_{2}+p_{2}x\label{eq:IC_lifetime_2}
\end{equation}
and
\begin{equation}
    F_{1,5}(x)=1-p_{1}+p_{1}x.\label{eq:IC_lifetime_3}
\end{equation}
The initial conditions in Eqs.~\cref{eq:IC_lifetime_1}, \cref{eq:IC_lifetime_2} and \cref{eq:IC_lifetime_3} are derived from the probability distribution of offspring in the next generation for type-1, type-2 and type-5 motifs respectively. For example, we get Eq.~\cref{eq:IC_lifetime_1} because there are no offspring with probability $(1-p_{1})^{2}$, one offspring with probability $2p_{1}(1-p_{1})$ and two offspring with probability $p_{1}^{2}$. In the initial condition, each of these probabilities is multiplied by $x^{0}$, $x^{1}$ and $x^{2}$ respectively, giving us $\mathbb{E}\left[X^{Z_{1,1}}\right]$ which is the pgf for the number of particles in generation 1 of a type-1 subtree $F_{1,1}(x)$. To find $\Omega_{k}$, the probability that a cascade has lifetime $k$, we use the property of pgfs that the probability that there are zero particles in generation $n$ is $\tilde{F}_{n}(0)$ and thus
\begin{equation}
    \Omega_{n}=\tilde{F}_{n}(0)-\tilde{F}_{n-1}(0).
\end{equation}
In \cref{fig:cascade_size_dist} (b), we show the distribution of cascade lifetimes for a doubly-Poisson distributed network comparing this theory to simulations.

\subsection{Joint Distributions\label{sec:joint_dist}}

Many interesting and practically-relevant properties of cascades depend on joint distributions of several quantities, and so it is important to move beyond univariate pgfs (as in \cref{sec:size_dist}) to develop methods for inverting multivariate pgfs. For example, we may wish to calculate the expected average tree depth (EATD), which we can find from the joint distribution of cascade size and cumulative depth. The \textit{cumulative depth} of a cascade is the sum of the depths of all of the nodes in the cascade. The average tree depth is the average depth of a node in the tree and the EATD is the expectation of this quantity over the ensemble of cascades. In a cascade, the average tree depth is given by the relationship
\begin{equation}
    \text{average tree depth} =\frac{\sum_{i\in V}\text{depth of node }i}{|V|}= \frac{\text{cumulative depth}}{\text{cascade size}},
\end{equation}
where $V$ is the set of all nodes in the cascade and $|V|$ is the size of the set $V$; i.e., the number of nodes in the cascade. To calculate the EATD, we need the joint distribution of cumulative depth and cascade size. We denote by $\tilde{X}_{n}$ the size of a cascade after $n$ generations and by $\tilde{Y}_{n}$ the cumulative depth of a cascade after $n$ generations. The joint pgf of cascade size and cumulative depth is
\begin{equation}
    \tilde{H}_{n}(x,y) = \mathbb{E}\left[x^{\tilde{X}_{n}}y^{\tilde{Y}_{n}}\right].
\end{equation}
We can write cascade size $\tilde{X}_{n}$ and cumulative depth $\tilde{Y}_{n}$ after $n$ generations in terms of the size and cumulative depth of the types 1 and 5 subtrees. A type-$i$ subtree is a subtree seeded by a type-$i$ motif, all six motif types are shown in \cref{fig:motifs_subtrees} (a). The sizes are denoted by ${X}_{n,1}$ and ${X}_{n,5}$ respectively and the depths are denoted by ${Y}_{n,1}$ and ${Y}_{n,5}$ respectively. In terms of subtree sizes, the cascade size is
\begin{equation}
    \tilde{X}_{n} = 1 + \sum_{i=1}^{t}X_{n,1}^{(i)} + \sum_{j=1}^{s}X_{n,5}^{(j)},
\end{equation}
where the seed node is in $t$ triangles and has $s$ single links and the cumulative depth is
\begin{equation}
    \tilde{Y}_{n} = \sum_{i=1}^{t}Y_{n,1}^{(i)} + \sum_{j=1}^{s}Y_{n,5}^{(j)}.
\end{equation}
This allows us to rewrite the pgf in terms of the joint pgfs of its type-1 and type-5 subtrees;
\begin{align}
    \tilde{H}_{n}(x,y) &=\mathbb{E}\left[ x^{1+\sum_{i} X_{n,1}^{(i)}+\sum_{j}\label{eq:jt_eq1} X_{n,5}^{(j)}}y^{\sum_{i} Y_{n,1}^{(i)}+\sum_{j} Y_{n,5}^{(j)}}\right]\\
    &= \sum_{s,t}\pi_{st}\mathbb{E}\left[ x^{1+\sum_{i=1}^{t} X_{n,1}^{(i)}+\sum_{j=1}^{s} X_{n,5}^{(j)}}y^{\sum_{i=1}^{t} Y_{n,1}^{(i)}+\sum_{j=1}^{s} Y_{n,5}^{(j)}}\middle|\text{ seed in }t\text{ triangles, }s\text{ links }\right]\nonumber\\
    &= x\sum_{s,t}\pi_{st}\mathbb{E}\left[x^{X_{n,5}}y^{Y_{n,5}}\right]^{s}\mathbb{E}\left[x^{X_{n,1}}y^{Y_{n,1}}\right]^{t}\nonumber\\
    &= x\tilde{f}\left( H_{n,5}(x,y), H_{n,1}(x,y)\right),\nonumber
\end{align}
where $\tilde{f}(x,y)$ is the pgf for the joint Newman-Miller distribution of the network and $H_{n,k}(x,y)$ is the pgf for the joint distribution of cascade size and cumulative depth for a subtree of type $k$ allowed to grow for $n$ generations. When calculating the cascade size of a type-$k$ subtree, $X_{n,k}$, similarly to in \cref{sec:size_dist}, we do not count the active or removed nodes in the seed motif because this would lead to counting some nodes multiple times when we consider ensembles of subtrees in the overall size. To calculate the full pgf, we will need expressions for $H_{n,1}(x,y)$, $H_{n,2}(x,y)$ and $H_{n,5}(x,y)$, using a method similar to that for the subtree size pgfs in \cref{sec:size_dist} we get,
\begin{align}
    H_{n,1}(x,y) &= \mathbb{E}\left[x^{X_{n,1}}y^{Y_{n,1}}\right]\\
    &= (1-p_{1})^{2}+2p_{1}(1-p_{1})xyH_{n-1,2}(xy,y)f_{r}\left(H_{n-1,5}(xy,y),H_{n-1,1}(xy,y)\right)\nonumber\\
    &+ p_{1}^{2}x^{2}y^{2}\left(f_{r}\left(H_{n-1,5}(xy,y),H_{n-1,1}(xy,y)\right)\right)^{2},\nonumber
\end{align}
\begin{align}
        H_{n,2}(x,y)&=\mathbb{E}\left[x^{X_{n,2}}y^{Y_{n,2}}\right]\\
        &=1-p_{2}+p_{2}xyf_{r}\left(H_{n-1,5}(xy,y),H_{n-1,1}(xy,y)\right)\nonumber
\end{align}
and
\begin{align}
        H_{n,5}(x,y) &=\mathbb{E}\left[x^{X_{n,5}}y^{Y_{n,5}}\right]\label{eq:jt_last_eq}\\
        &=1-p_{1}+p_{1}xyf_{q}\left(H_{n-1,5}(xy,y),H_{n-1,1}(xy,y)\right),\nonumber
\end{align}
where $f_{r}$ and $f_{q}$ are the pgfs for the excess Newman-Miller distribution from a triangle and a single link respectively, these are given in Eqs.~\cref{eq:fr} and \cref{eq:fq}. We can find the full joint pgf of cascade size and cumulative depth by iterating through these equations from $n=0$ with initial conditions $H_{0,1}(x,y)=H_{0,2}(x,y)=H_{0,5}(x,y)=1$. To find the joint probability distribution from the joint pgf we use an inverse fast Fourier transform method which we describe next in \cref{section:1D_inversion}. While methods that use computer algebra systems have been used in \cite{brummit2012}, we are not aware of other work on cascade dynamics that uses this method to recover the joint probability distribution of cascade properties. The main issue with using computer algebra systems to calculate the probabilities from pgfs is that they do not allow us to symbolically calculate the pgf for a large number of generations. This limits us in the number of sizes that we can calculate the probabilities for and also reducing the accuracy in the approximation in the limit as the number of generations $n\to\infty$. This has motivated us to propose the inverse fast Fourier method of recovering the probability distribution which we describe next in \cref{section:1D_inversion}, which does not have the same shortcomings as the aforementioned methods.

\section{Inverting multivariate pgfs to recover probabilities\label{section:1D_inversion}}

Given a pgf (or at least an iterative method for calculating it), in many cases, we would like to access the probability distribution of the quantity that it describes. For one-dimensional pgfs, in the network dynamics literature, the most common inversion method used is that of Cavers \cite{cavers1978}. In Cavers' method, a link between the pgf and the z transform is utilised and the inversion is expressed in the form of an inverse z transform. The inverse z transform requires the computation of a contour integral in the complex plane, which is approximated using the trapezoidal rule. The trapezoidal-rule approximation is equivalent to taking the inverse discrete Fourier transform of the pgf evaluated at $K$ evenly spaced points around the unit circle centred at zero on the complex plane. In practice, we use the inverse fast Fourier transform, which is a fast, efficient algorithm for computing inverses of discrete Fourier transforms, in the place of the inverse discrete Fourier transform. Here, we introduce an alternative derivation which more naturally extends to the inversion of higher-dimensional pgfs. Our method has the same computational implementation as Cavers' method in one dimension. Given the pgf
\begin{equation}
    B(z)=\sum_{i=0}^{\infty}p_{i}z^{i},
\end{equation}
we aim to find the probabilities $\{p_{i}\},~i\in\{0,1,2,...\}$, where $p_{i}$ is the probability that the random variable, with probability distribution described by the pgf $B(z)$, is equal to $i$. The discrete Fourier transform of the first $K$ probabilities is denoted by the set $\{P_{l}\}$, $l\in\{0,1,...,K-1\}$, and is given by the relationship
\begin{equation}
    P_{l} = \sum_{k=0}^{K-1}p_{k}(e^{\frac{-2\pi i}{K}l})^{k}\approx B(e^{\frac{-2\pi i}{K}l}),\label{eq:1d_inversion}
\end{equation}
which is approximately the pgf $B(z)$ evaluated at $e^{\frac{-2\pi i}{K}l}$ for $K$ sufficiently large and $l\in\{0,1,2,...,K-1\}$ if we assume that $p_{k}\to 0$ as $k\to \infty$. In our examples, we let $K=100$. For example, to evaluate the pgf for cascade size, we iterate through Equations \cref{eq:size_pgf_full} to \cref{eq:size_pgf_6} setting $x=e^{\frac{-2\pi i}{K}l}$ and do this for each $l$, this gives us a vector holding values of the approximation of the discrete Fourier transform of the cascade size probabilities. Once we have evaluated the pgf at each of these points, we use the inverse fast Fourier transform on that set of transformed points to obtain the an approximation for the first $K$ probabilities from the pgf, the larger a value for $K$ that we choose, the closer the approximation will be to the true values. In \cref{fig:cascade_size_dist} (a) we show the theoretical distribution of cascade size which is found using this method to invert the pgf for cascade size that we described in \cref{sec:size_dist}.
\par
This method naturally extends to two dimensions (and beyond), allowing us to retrieve, for example, the joint probability distribution of cascade size and cumulative depth from \cref{sec:joint_dist}. The joint probability distribution of random variables $X_{1}$ and $X_{2}$ is given by the set of probabilities $\{p_{k_{1},k_{2}}\}$ where $p_{k_{1},k_{2}}$ is the probability that $X_{1}=k_{1}$ and $X_{2}=k_{2}$. The general joint pgf has the form
\begin{equation}
    A(x,y) = \sum_{k_{1}=0}^{\infty}\sum_{k_{2}=0}^{\infty}p_{k_{1},k_{2}}x^{k_{1}}y^{k_{2}}.
\end{equation}
and similar to the univariate case, if we evaluate the pgf for all combinations of $x\in\{1$, $e^{-\nicefrac{2\pi i}{K_{1}}}$, $e^{-\nicefrac{4\pi i}{K_{1}}}$, $...$, $e^{-\nicefrac{2(K_{1}-1)\pi i}{K_{1}}}\}$ and $y\in\{1$, $e^{-\nicefrac{2\pi i}{K_{2}}}$, $e^{-\nicefrac{4\pi i}{K_{2}}}$, $...$, $e^{-\nicefrac{2(K_{2}-1)\pi i}{K_{2}}}\}$, we get the 2-dimensional discrete Fourier transform of the joint probabilities of all combinations of the first $K_{1}$ values for $X_{1}$ and the first $K_{2}$ values for $X_{2}$; i.e., $k_{1}$ varying from 0 to $K_{1}-1$ and $k_{2}$ varying from 0 to $K_{2}-1$;
\begin{equation}
    P_{l,m} = \sum_{k_{1}=0}^{K_{1}-1} \sum_{k_{2}=0}^{K_{2}-1} p_{k_{1}, k_{2}}\left(e^{-\frac{2\pi i l}{K_{1}}}\right)^{k_{1}}\left(e^{-\frac{2\pi i m}{K_{2}}}\right)^{k_{2}}\approx A\left(e^{-\frac{2\pi i}{K_{1}}l}, e^{-\frac{2\pi i}{K_{2}}m}\right).\label{eq:2dfft}
\end{equation}
for $K_{1}$ and $K_{2}$ sufficiently large. In the case of the joint pgf for cascade size and cumulative depth, this means iterating through Equations \cref{eq:jt_eq1} to \cref{eq:jt_last_eq} with $x=e^{-\nicefrac{2\pi i l}{K_{1}}}$ and $y=e^{-\nicefrac{2\pi i m}{K_{2}}}$ for all $l,m$ combinations, $l\in\{0,1,...,K_{1}-1\}$ and $m\in\{0,1,...,K_{2}-1\}$. From Eq.~\cref{eq:2dfft} if we evaluate the pgf for all combinations of $x\in\{1$, $e^{-\nicefrac{2\pi i}{K_{1}}}$, $e^{-\nicefrac{4\pi i}{K_{1}}}$, $ ...$,$e^{-\nicefrac{2(K_{1}-1)\pi i}{K_{1}}}\}$ and $y\in\{1,e^{-\nicefrac{2\pi i}{K_{2}}}$, $ e^{-\nicefrac{4\pi i}{K_{2}}}$, $...$, $e^{-\nicefrac{2(K_{2}-1)\pi i}{K_{2}}}\}$ and take the 2-dimensional inverse fast Fourier transform, we will recover an approximation for the joint probabilities $p_{k_{1},k_{2}}$ for  all combinations of the first $K_{1}$ values for $k_{1}$ and the first $K_{2}$ values for $k_{2}$. As for the uni-variate case, the accuracy in the approximation increases as $K_{1}$ and $K_{2}$ increase. If we use this method to recover the joint distribution for cascade size and cumulative depth from the joint pgf derived in \cref{sec:joint_dist}, for example, we can calculate theoretical properties of the cascades including, but not limited to, the EATD and the correlation between cascade size and cumulative depth that we will describe in \cref{sec:EATD_correlation}.\par
In $N$ dimensions the pgf would take the form,
\begin{equation}
    F(x_{1},x_{2},...,x_{N}) = \sum_{k_{1}=0}^{\infty}\sum_{k_{2}=0}^{\infty}...\sum_{k_{N}=0}^{\infty}p_{k_{1},k_{2},...,k_{N}}x_{1}^{k_{1}}x_{2}^{k_{2}}...x_{N}^{k_{N}},
\end{equation}
which, when evaluated for each combination of $x_{i} \in\{1,e^{-\nicefrac{2\pi i}{K_{i}}}, e^{-\nicefrac{4\pi i}{K_{i}}},...,e^{-\nicefrac{2(K_{i}-1)\pi i}{K_{i}}}\}$, $i\in\{1,2,...,N\}$ becomes approximately the $N$-dimensional discrete Fourier transform of the probabilities $p_{k_{1},k_{2},...,k_{N}}$. As for the univariate and bivariate cases, taking the inverse fast Fourier transform gives us an approximation to the $N$-dimensional joint distribution.

\subsection{Pearson correlation coefficient and EATD\label{sec:EATD_correlation}}

In this subsection, we show how the Pearson correlation coefficient between cascade size and cumulative depth and the EATD are calculated from the joint probability distribution that we described in \cref{sec:joint_dist}. The joint distribution is recovered from its pgf in the way that we have just shown. The Pearson correlation coefficient can be expressed in terms of the moments of the distribution
\begin{equation}
    \rho_{\tilde{X},\tilde{Y}} = \frac{\mathbb{E}[\tilde{X}\tilde{Y}]-\mathbb{E}[\tilde{X}]\mathbb{E}[\tilde{Y}]}{\sqrt{\mathbb{E}[\tilde{X}^{2}]-\mathbb{E}[\tilde{X}]^{2}}\sqrt{\mathbb{E}[\tilde{Y}^{2}]-\mathbb{E}[\tilde{Y}]^{2}}}
\end{equation}
these moments can be calculated from the probability distribution. For unclustered networks, Gleeson et al.~\cite{gleeson2021} described a method for calculating the EATD using an integral formula; however, their method does not extend to clustered networks. An alternative method of calculating the EATD is to use the joint probability distribution of cascade size and cumulative depth, this method is not restricted to unclustered networks. We compute the EATD using the formula
\begin{equation}
    \text{EATD} = \sum_{i,j=0}^{\infty}P\left[\text{cascade size }=i\text{ \& cumulative depth }=j\right]\frac{j}{i}.
\end{equation}
In \cref{tab:EATD_correlation}, we show the EATD and Pearson's correlation found using this method and compare to Monte-Carlo simulations on two synthetic networks and three real-world networks.

\begin{table}[tbhp] 
\footnotesize \caption{Table showing both the EATD and Pearson correlation, $\rho$, from the MTBP theory and from 1 million Monte-Carlo simulations. We show results for a doubly-Poisson Newman-Miller network ($\mu=1$, $\nu=4$) with 10,000 nodes and for two empirical networks; the power-grid network \cite{watts1998collective}, the largest connected component of the science co-authorship network \cite{Newman2006} and the C.~elegans metabolic network \cite{duch2005}. For the empirical networks, in the calculation of the theoretical EATD and $\rho$, we estimated the Newman-Miller distribution ($\pi_{st}$ from \cref{sec:newman_miller}) using the edge-disjoint edge clique cover (EECC) described in \cref{sec:EECC_algorithm}.\label{tab:EATD_correlation}}
\begin{center}
\begin{tabular}{|c|c|c|c|c|c|c|}
\hline Network&$p_{1}$&$\alpha$&simulation EATD&theory: EATD& simulation $\rho$& theory: $\rho$\\\hline
Newman-Miller&0.05&0&0.347&0.333&0.905&0.898\\
Newman-Miller&0.02&0.2&0.126&0.124&0.925&0.932\\
power grid&0.04&0.15&0.062&0.059&0.796&0.863\\
science co-authorship&0.01&0.1&0.041&0.027&0.938&0.960\\
C.~elegans&0.005&0.005&0.007&0.026&0.976&0.935\\\hline
\end{tabular}
\end{center}
\end{table}

\section{Application to real-world networks\label{sec:app_to_data}}

To apply these methods to empirical networks, we need to recover the joint distribution of triangles and single links from the network, this allows us to find the pgf for the distribution $\tilde{f}(x,y)$ and for the excess distributions $f_{r}(x,y)$ and $f_{q}(x,y)$, as given in \cref{sec:newman_miller}, for a Newman-Miller network with the same joint distribution as the empirical network. An assumption of the MTBP method for modelling complex contagion on clustered networks is that the cliques do not share edges --- they are edge-disjoint or, equivalently, are connected in a locally tree-like manner --- in practice, many real-world networks have cliques which are not edge-disjoint. In Burgio et al.~\cite{burgio21} they propose the Edge-Disjoint Edge Clique Cover (EECC), a cover which imposes, as the name suggests, the constraint that the cliques are edge-disjoint. With the EECC, we can approximate $\tilde{f}$, and thus $f_{q}$ and $f_{r}$.
\par
\begin{figure}[h]
    \centering
\includegraphics[width=0.8\textwidth]{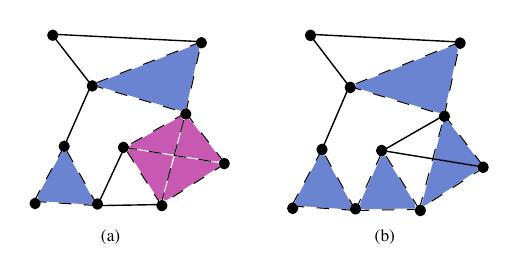}
    \caption{(a) An edge-disjoint edge clique cover (EECC) of a network (black nodes and edges) showing the cliques included in the cover, the 3-cliques shaded in blue and 4-clique shaded in pink are included in the EECC, the edges that do not form part of these cliques are included in the EECC as 2-cliques. The dashed edges represent edges in a higher-order clique and the solid edges connect two nodes in a 2-clique. (b) The EECC of the same network when we restrict the maximum clique size to be 3.\label{fig:eecc}}
\end{figure}

In our examples here, we assume that the network is composed of triangles and links but not higher-order cliques. Since the cliques in real-world networks can have more than three nodes; i.e., a triangle, we approximate any higher-order cliques as a combination of triangles and links, as shown in \cref{fig:eecc} (b). \Cref{fig:eecc} (a) shows the decomposition of the network into edge-disjoint cliques where there is no restriction on the size of the cliques, this is the EECC introduced by Burgio et al.~\cite{burgio21}. In \cref{sec:EECC_algorithm} we present our alteration of the EECC algorithm introduced by Burgio et al.~\cite{burgio21}, which imposes that the cliques in the EECC have maximum size 3; i.e., are at most triangles, this is the algorithm that we use here. This decision was due to the fact that the equations that we set up in previous sections assume that the network is fully composed of triangles and single links; in theory, if these systems of equations are extended to network distributions with higher-order cliques then it would be natural to extend the algorithm to capture these larger cliques. However, it is possible that extending the systems of equations beyond relatively small clique sizes would be prohibitively tedious by hand and an extension to higher order cliques is likely to require the derivation of the equations to be automated.

\subsection{Implementation of the EECC}
\begin{figure}[h!]
    \centering
         \begin{subfigure}{0.45\textwidth}
         \centering
         \includegraphics[width=\textwidth]{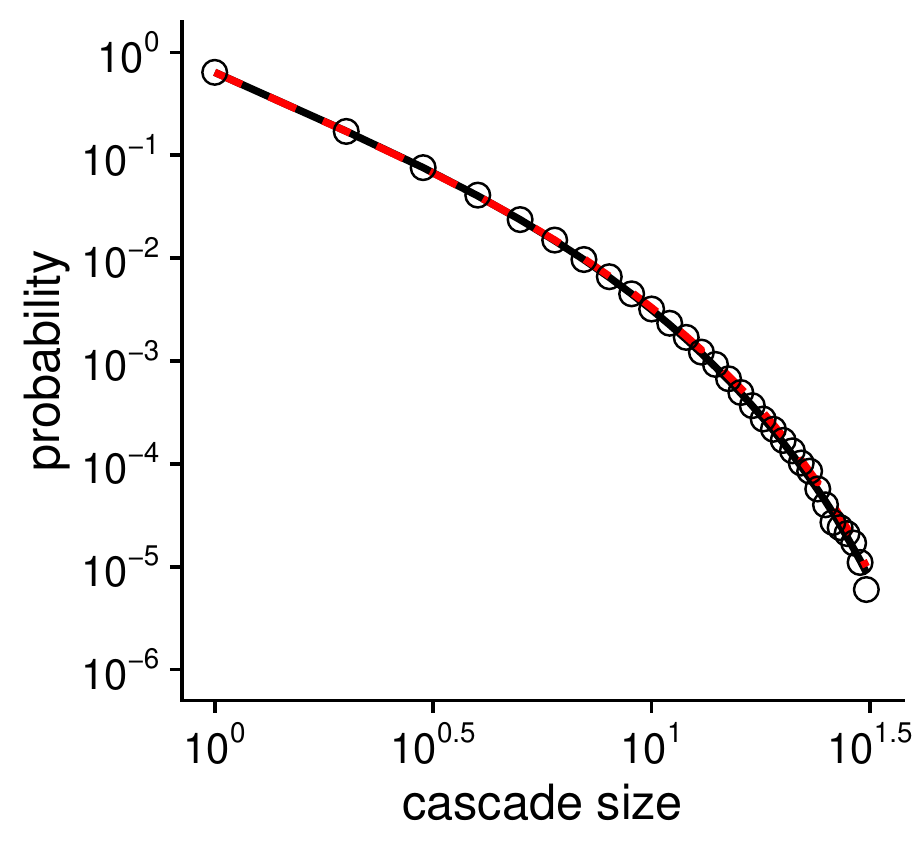}
         \caption{pmf: $p_{1}=0.05$ and $\alpha=0$}
     \end{subfigure}
     \hfill
     \begin{subfigure}{0.45\textwidth}
         \centering
         \includegraphics[width=\textwidth]{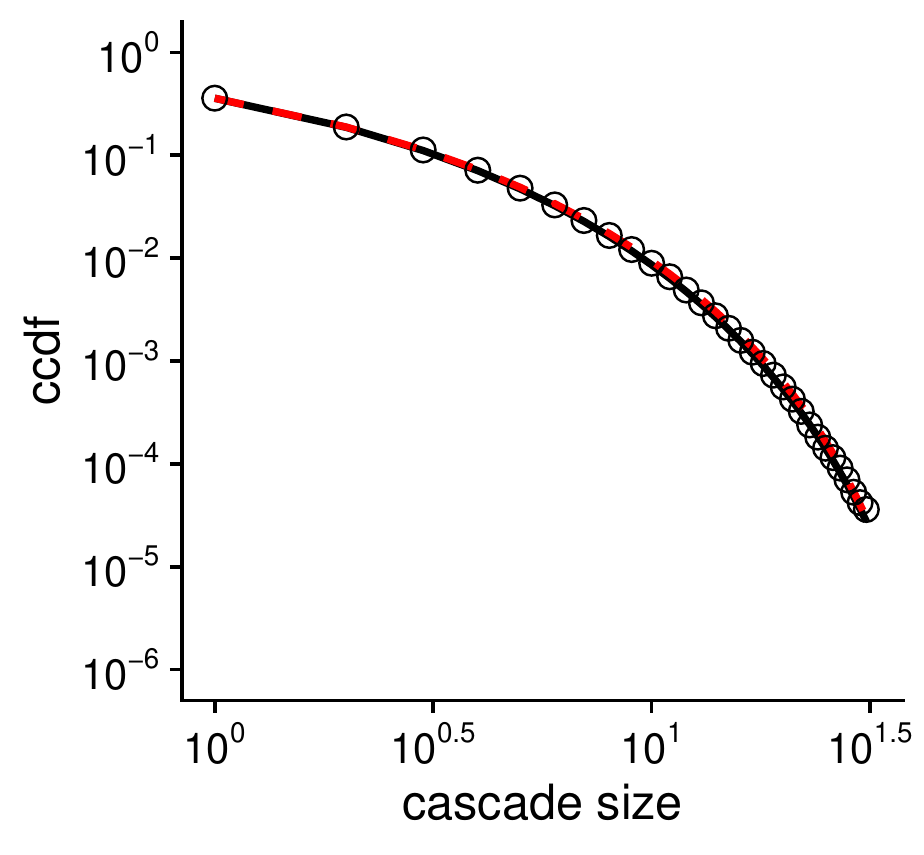}
         \caption{ccdf: $p_{1}=0.05$ and $\alpha=0$}
     \end{subfigure}
         \begin{subfigure}{0.45\textwidth}
         \centering
         \includegraphics[width=\textwidth]{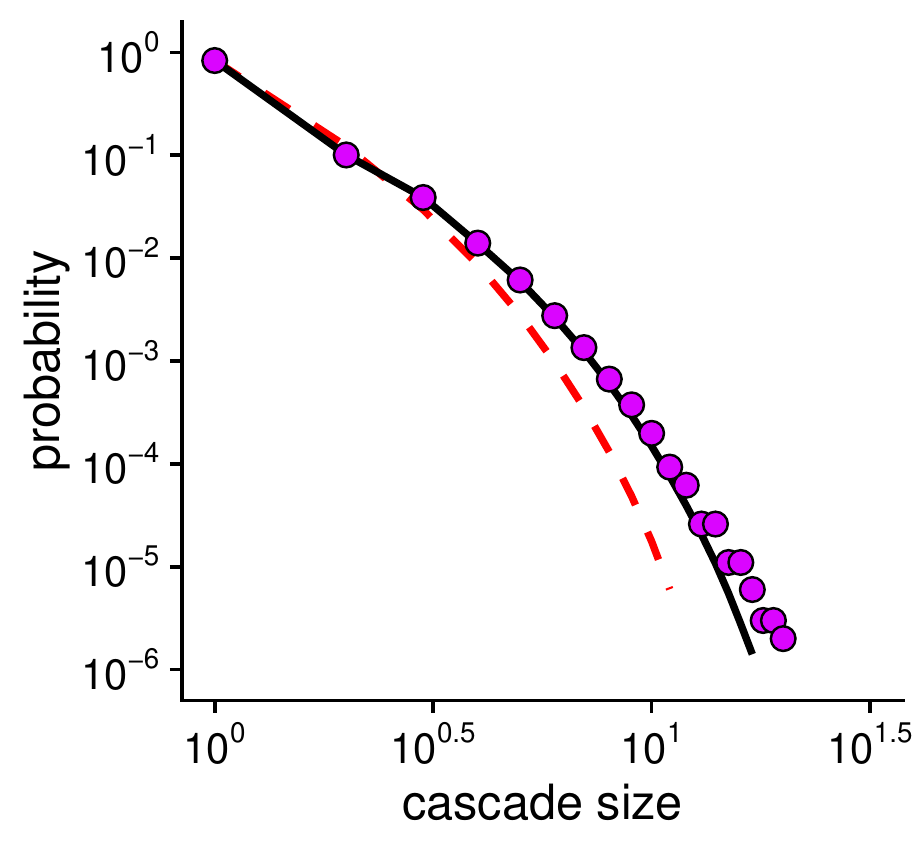}
         \caption{pmf: $p_{1}=0.02$ and $\alpha= 0.2$}
     \end{subfigure}
     \hfill
     \begin{subfigure}{0.45\textwidth}
         \centering
         \includegraphics[width=\textwidth]{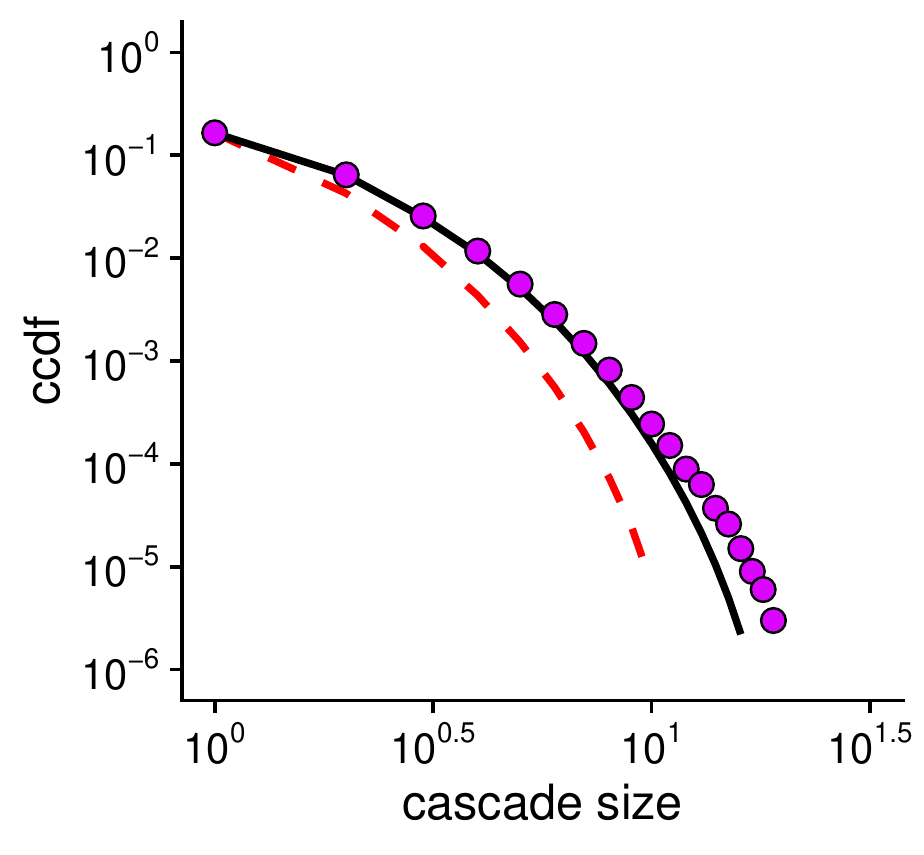}
         \caption{ccdf: $p_{1}=0.02$ and $\alpha= 0.2$}
     \end{subfigure}

    \caption{The pmfs (probability distributions) (a) and (c) and ccdfs (b) and (d) for cascade size for a synthetically generated network of size 5,000 which follows a doubly-Poisson distribution with parameters $\mu=1$ and $\nu=4$ for the dynamics described in \cref{sec:adoption_dynamics}. In (a) and (b) the complex contagion parameters are $p_{1} = 0.05$ and $\alpha=0$, this is a simple contagion, in (c) and (d) the parameters are $p_{1} = 0.02$ and $\alpha=0.2$, this is a complex contagion. The points are from simulating 1 million cascades on the network, the solid line is from estimating the pgf of the link-triangle distribution from the network using an EECC and the dashed line is a simple branching process approximation of the dynamics. In (a) and (b) we have removed the colour from the points to make the curves from theory more visable.\label{fig:EECC_comparison_synthetic}}
\end{figure}

In this subsection, we apply the EECC to a synthetic network generated from the doubly-Poisson distribution and compare the results to what would be attained from a single-edge decomposition (SED), which assumes that the network is fully composed of single edges and thus the spreading dynamics constitute a simple branching process. In \cref{fig:EECC_comparison_synthetic} we show the results of using the EECC on a synthetically generated network with a doubly-Poisson distribution with parameters $\mu = 1$ and $\nu=4$ (mean degree 9). In this network, the assumptions of the MTBP theory are very close to being met; i.e., the cliques do not overlap and the network is very large (5000 nodes). We applied the EECC to the network to approximate its joint triangle-link distribution and use this to find the distribution of cascade sizes according to the method described in \cref{sec:size_dist}. We compare the results of this approach to using the SED and the method for calculating the cascade size distribution given in \cite{gleeson2021}, for completeness, this method is summarised in \cref{sec:SED_cascade_size}.
\par
In \cref{fig:EECC_comparison_synthetic} (a) and (b) we examine the distribution of cascade sizes for simple contagion dynamics where $p_{1}=0.05$ and $\alpha=0$. For a simple contagion on this network, we see that the theoretical distribution of cascade size matches well to the simulations when we account for the presence of triangles in the network. However, when we compare the MTBP which accounts for the the presence of triangles to a simple branching process approximation (using the SED) which does not, there is minimal difference in the probability distributions, both curves match the simulations very well --- this result is not particularly surprising, Melnik et al.~\cite{melnik2011} found that tree-based theory for dynamical processes on various highly clustered networks yielded accurate results despite the locally tree-like assumption being clearly violated. In \cref{fig:EECC_comparison_synthetic} (c) and (d) we examine the cascade size distribution for complex-contagion dynamics where $p_{1}= 0.02$ and $\alpha=0.2$, we compare simulations on the network to the theoretical distribution of sizes and find very strong agreement between the theory and simulations. In the figures we include the simple branching process approximation of the same dynamics and see that accounting for the clustering in the network gives a more accurate theoretical result. When we compare the result for the theory that accounts for the presence of triangles to the theory that does not, we see that for these parameters not accounting for the presence of triangles leads to smaller cascades than are observed in simulations, again showing the importance of accounting for clustering when modelling complex contagion on clustered networks.

\subsection{Application of the EECC to real-world networks}

\begin{figure}[h!]
    \centering
         \begin{subfigure}{0.47\textwidth}
         \centering
         \includegraphics[width=\textwidth]{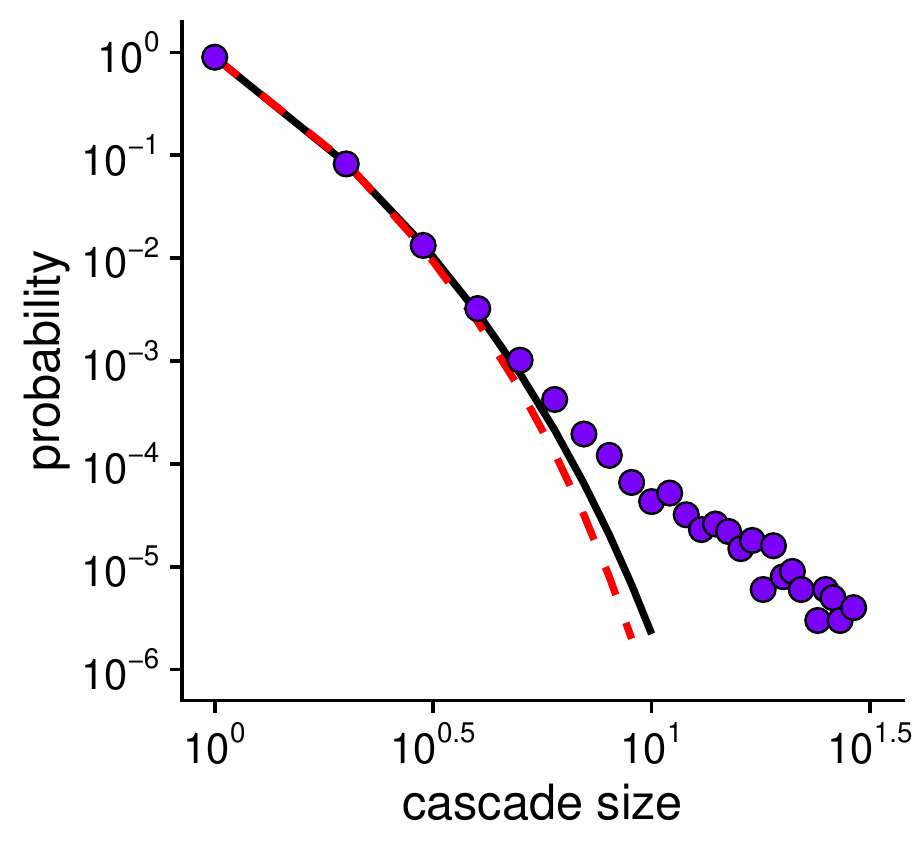}
         \caption{power grid: probability distribution}
     \end{subfigure}
     \hfill
     \begin{subfigure}{0.47\textwidth}
         \centering
         \includegraphics[width=\textwidth]{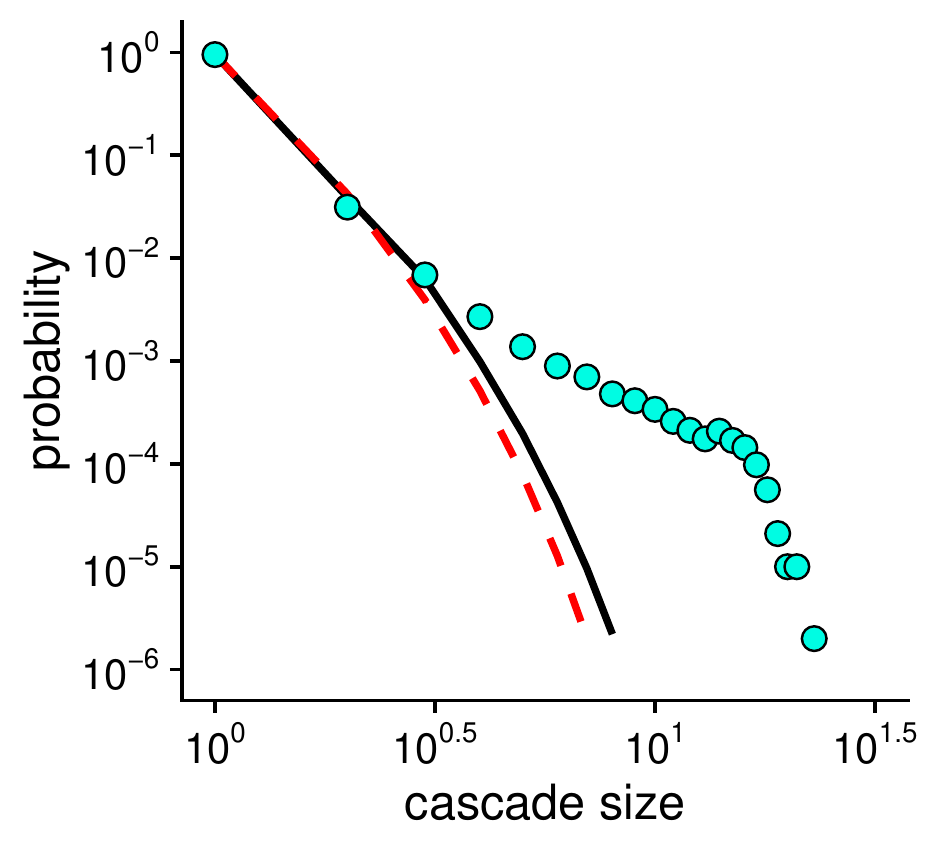}
         \caption{science co-authorship: probability distribution}
     \end{subfigure}
     \begin{subfigure}{0.47\textwidth}
         \centering
         \includegraphics[width=\textwidth]{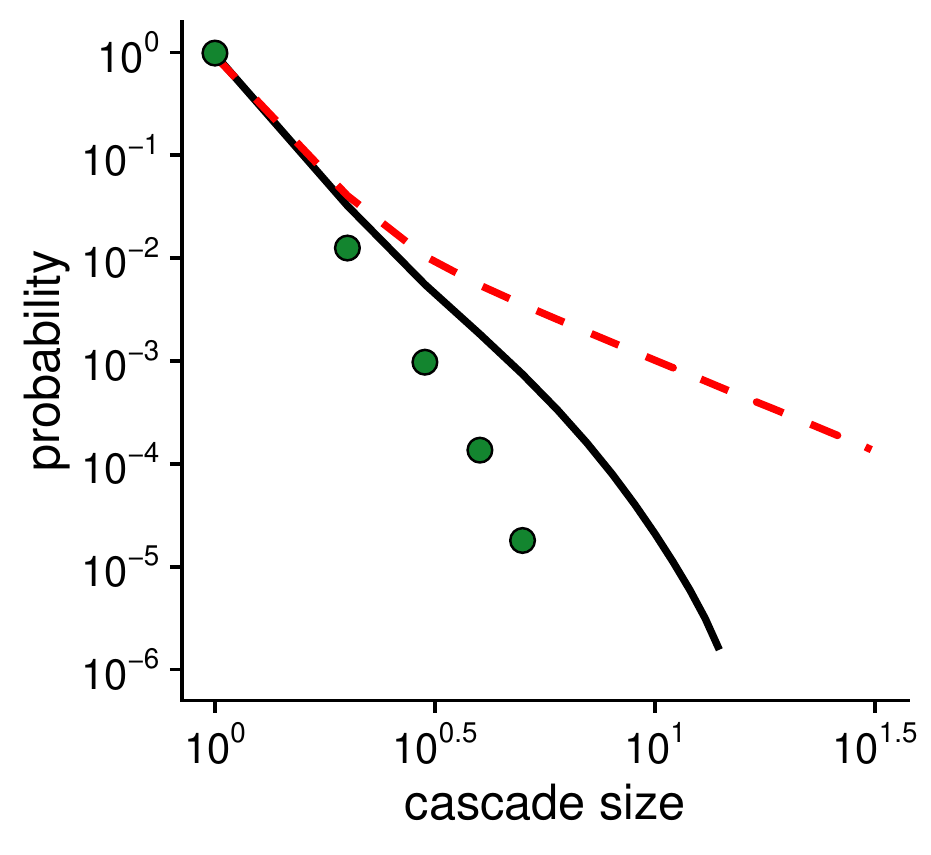}
         \caption{C.~elegans: probability distribution}
     \end{subfigure}
    \caption{The probability distributions for cascade size for complex contagion dynamics on (a) the power-grid network \cite{watts1998collective} ($p_{1}=0.04$ and $\alpha=0.15$), (b) the largest connected component of the science co-authorship network \cite{Newman2006} ($p_{1}=0.01$ and $\alpha=0.1$) and (c) the C. elegans metabolic network \cite{duch2005} ($p_{1}=0.005$ and $\alpha = 0.005$). In all subfigures, we compare the theoretical distribution for cascade size accounting for the presence of triangles (network distribution approximated using the EECC) (black solid line) and not accounting for the presence of triangles (red dashed line) to 1 million simulations (coloured points).\label{fig:real_world_pdfs}}
\end{figure}

\begin{figure}[h!]
    \centering
         \begin{subfigure}{0.45\textwidth}
         \centering
         \includegraphics[width=\textwidth]{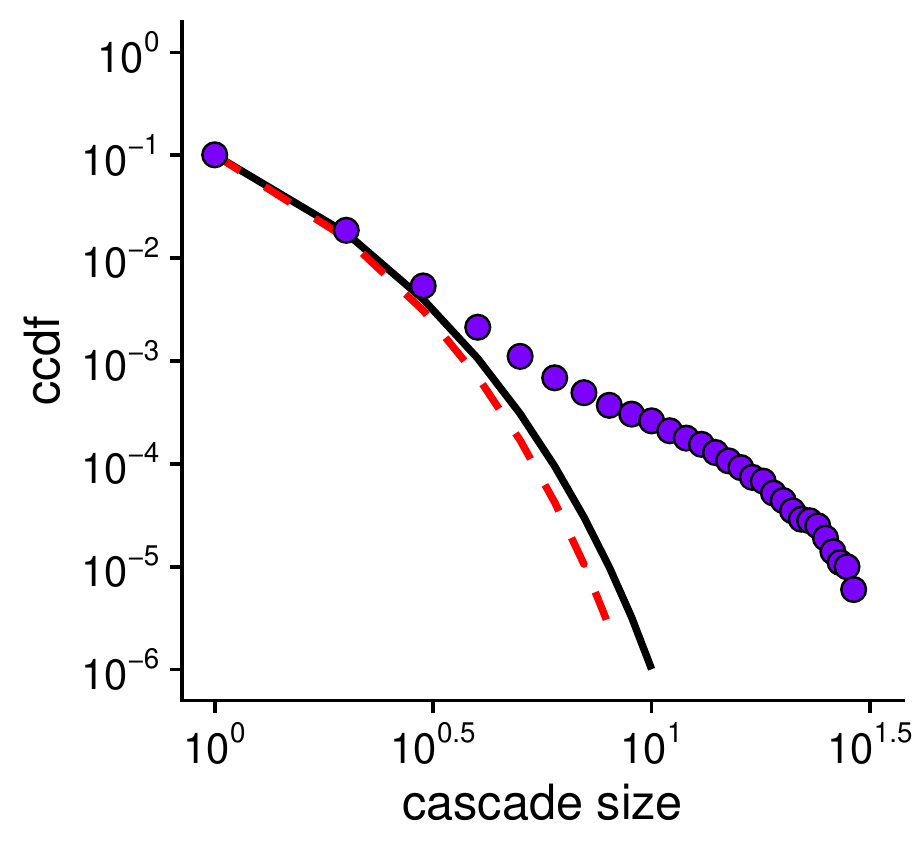}
         \caption{power grid: ccdf}
     \end{subfigure}
     \begin{subfigure}{0.45\textwidth}
         \centering
         \includegraphics[width=\textwidth]{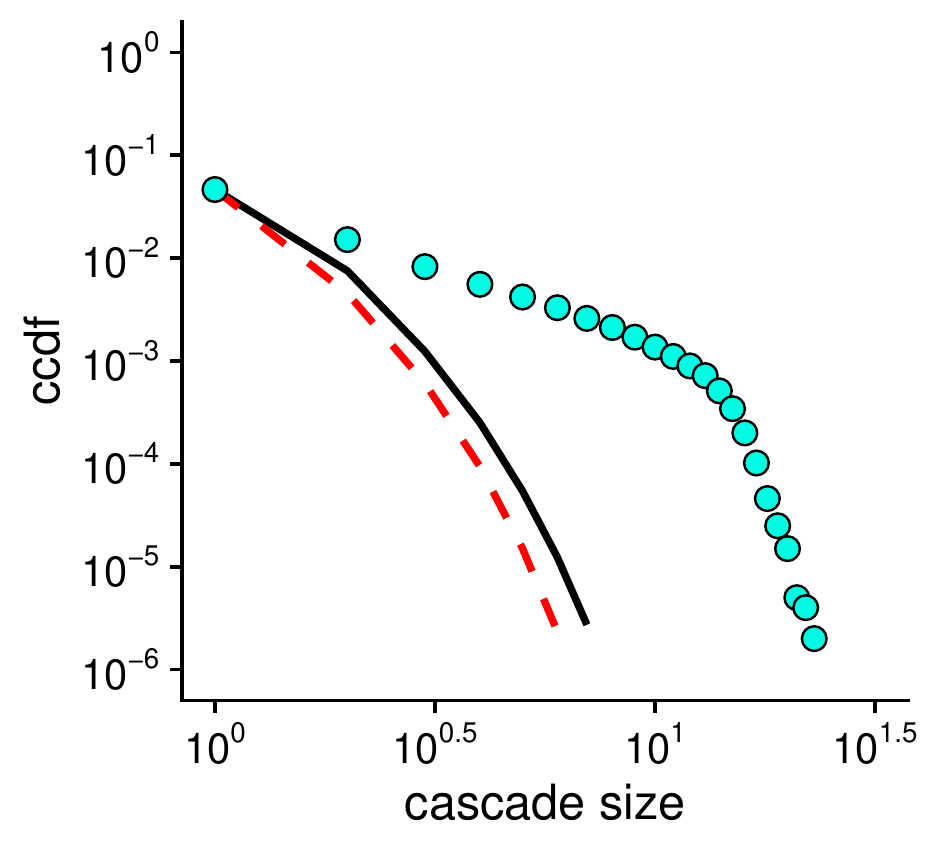}
         \caption{science co-authorship: ccdf}
     \end{subfigure}
     \begin{subfigure}{0.45\textwidth}
         \centering
         \includegraphics[width=\textwidth]{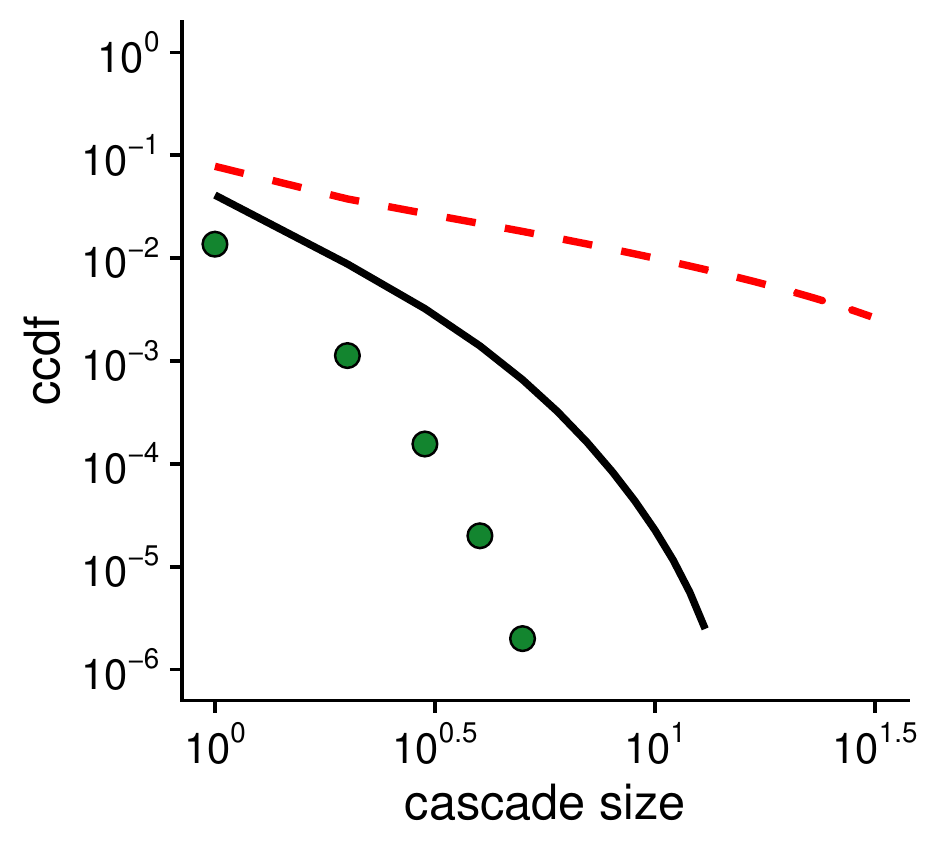}
         \caption{C.~elegans: ccdf}
     \end{subfigure}
     \caption{The complimentary cumulative distribution function (ccdf) for cascade size for complex contagion dynamics on (a) the power-grid network \cite{watts1998collective} ($p_{1}=0.04$ and $\alpha=0.15$), (b) the largest connected component of the science co-authorship network \cite{Newman2006} ($p_{1}=0.01$ and $\alpha=0.1$) and (c) the C. elegans metabolic network \cite{duch2005} ($p_{1}=0.005$ and $\alpha = 0.005$). In all subfigures, we compare the theoretical distribution for cascade size accounting for the presence of triangles (network distribution approximated using the EECC) (black solid line) and not accounting for the presence of triangles (red dashed line) to 1 million simulations (coloured points).\label{fig:real_world_ccdfs}}
\end{figure}

So far, we have shown that we can derive interesting statistics about complex contagion dynamics on clustered networks --- or at least networks which are structurally composed of cliques which are edge disjoint. While we remain in the (arguably unrealistic) regime where the maximum clique size is 3, we would like to know how these methods work when we apply them to real-world networks which may have correlations and clique size distributions which are not present in the synthetic distributions that we have worked with so far. We apply these methods to three well-studied real-world networks; the power-grid network \cite{watts1998collective} (4941 nodes, 6594 edges, $C_{\Delta} = 0.10$\footnote{The clustering coefficient, $C_{\Delta}$, is the ratio between the number of triangles and the number of connected triples in the network.}), the largest connected component of the science co-authorship network \cite{Newman2006} (379 nodes, 914 edges, $C_{\Delta}=0.43$) and the C.~elegans metabolic network \cite{duch2005} (453 nodes, 4596 edges, $C_{\Delta}=0.12$). We chose these networks because they are relatively large --- reducing the impact of finite-size effects --- but small enough that the EECC algorithm finds a cover in a reasonable run time. In \cref{fig:real_world_pdfs} (a) and \cref{fig:real_world_ccdfs} (a) we show the probability distribution and ccdf respectively of cascade size for the power-grid network with complex contagion parameters $p_{1}=0.04$ and $\alpha=0.15$, we compared 1 million simulations on the network to the theoretical distributions accounting for triangles with the MTBP and also not accounting for triangles using the SED. While the theory matches quite well to the simulated cascade size probability distribution, in \cref{fig:real_world_pdfs} (a) and \cref{fig:real_world_ccdfs} (a), for smaller cascade sizes, the difference in what is predicted by the SED and by the MTBP with the EECC is very small. The tree-like approximation made in the SED does quite well despite the power-grid network being clustered. In the tails of the cascade size distribution both of the theoretical distributions do not match the simulations very well, we speculate that this is due to the larger cliques present in the network which are not accounted for in the theoretical approach. Similarly, we applied the methods to the largest connected component of the science co-authorship network \cite{Newman2006}, using complex contagion parameters $p_{1}=0.01$ and $\alpha=0.1$, we show the probability distribution and ccdf in \cref{fig:real_world_pdfs} (b) and \cref{fig:real_world_ccdfs} (b). The EECC was also applied to this network in Ref.~\cite{Mann2022_thesis}. We found that while the simulations match the theory extremely well for cascade sizes up to size three, which accounts for $99.2\%$ of all cascades, our theory with the EECC under-predicts the number of large cascades that we observe in simulations. We anticipate that this is because there are larger cliques in the empirical network that are not accounted for when we use the EECC with maximum clique size 3 to approximate the clique distribution of the network. Given that this network is not very large (379 nodes) and that branching processes assume an infinitely large network, it is important to check that the network size is not causing the discrepancy between the theoretical distributions and the simulated cascades. In \cref{sec:finite_size}, we explore the impact of finite-size effects and find that these are minimal, again reinforcing our hypothesis that the difference between the theory and the simulations is caused by the theory not accounting for the higher-order cliques (with more than three nodes). Finally, we applied this method to the C.~elegans metabolic network \cite{duch2005} for complex contagion with parameters $p_{1}=0.005$ and $\alpha=0.005$. For this network, the simple branching process theory poorly approximates the cascade-size distribution (probability distribution and ccdf) as shown by the red dashed line in \cref{fig:real_world_pdfs} (c) and \cref{fig:real_world_ccdfs} (c) respectively. When we account for the presence of triangles (as shown by the black solid line) using the MTBP with the EECC to approximate the Newman-Miller distribution, the theory comes much closer to the simulations on the empirical network which suggests to us that it is the presence of higher-order structures in the C.~elegans metabolic network that leads the simple branching process to poorly describe the dynamics. By accounting for links and triangles only, the MTBP substantially improves the theoretical distribution of cascade size, when we compare to simulations on the C.~elegans metabolic network.
\par

\section{Discussion and Conclusions\label{sec:discussion}}

In this paper, we have extended the multi-type branching process (MTBP) theory originally proposed in Ref.~\cite{2022Keating} in three main directions; first, the clustered networks studied in \cite{2022Keating} were highly stylised and we showed how the more general Newman-Miller class of networks can be used to incorporate clustering into the branching process framework; then, we used the MTBP to derive more results, the focus here was on looking at full distributions of cascade sizes, cascade lifetimes and the joint distribution of cascade size and cumulative depth which enabled us to calculate the expected average tree depth and the correlation between cascade size and cumulative depth, this contrasts with a focus on the average behaviour of the cascades in our previous paper \cite{2022Keating}. Finally, we showed how to apply the theory to real-world networks. To apply the theory to real-world networks, we approximated the real-world network structure as a Newman-Miller network using the edge-disjoint edge clique cover (EECC) proposed by Burgio et al.~\cite{burgio21}. A second contribution of this paper is our derivation of the numerical inversion method for multi-variate probability generating functions (pgfs). While Cavers' method \cite{cavers1978} has been extended to bivariate pgfs \cite{Antal2010}, our alternative derivation has the advantage of extending naturally to pgfs of any dimension.
\par
In Keating et al.~\cite{2022Keating}, we focused on a very stylised model of network clustering for ease of explanation. These networks were infinitely large and we assumed that the local structure of the network was invariant throughout the network; e.g., we looked at networks where every node was in two 4-cliques and a network where every node was in four 2-cliques and one 3-clique. Here, we showed how the method can be extended to incorporate any Newman-Miller network distribution describing the distribution of triangle and single-link membership of the nodes in a network. In our examples, we use the doubly-Poisson distribution as used by Newman \cite{Newman2009}. While we did not delve into it here, using the probability generating functions for the Newman-Miller network distributions, all of the results on the average behaviour of the cascades from Ref.~\cite{2022Keating} can be extended to this more general class of networks. For example, we can calculate the expectations from the offspring distribution pgfs and use this to find the elements of the mean matrix, allowing us to find the cascade condition and the expected cascade size using the methods described in Ref.~\cite{2022Keating}. Furthermore, while we focused here on Newman-Miller networks, which have a maximum clique size of three, in principle the method can be extended beyond this restriction to multivariate distributions of cliques of any size, this would mean that we would need to incorporate different \textit{types} into the MTBP for which the transition probabilities may be calculated either by hand or using a computer algorithm. To incorporate very large cliques into the MTBP, we expect that it will not be feasible to do this by hand and it would be necessary to write a computer algorithm to do this for us --- we did not attempt to do this as part of this work.
\par
Gleeson et al.~\cite{gleeson2021} successfully used simple branching processes to model information cascades on Twitter. They analytically derived full distributions of cascade lifetimes and cascade size, and calculated the expected average tree depth (EATD) and structural virality. For complex-contagion dynamics on clustered networks, we here showed how, using pgfs, we can find the cascade size distribution, distribution of cascade lifetimes and the EATD. Our method for calculating the cascade size distribution, as in Ref.~\cite{gleeson2021}, involves finding the pgf for the cascade size distribution and recovering the probability distribution using an inverse fast Fourier transform method. In \cref{section:1D_inversion}, we explained how inverse fast Fourier transforms can be used to recover an accurate approximation of the probabilities, this naturally extends to higher-dimensional pgfs. We use inverse fast Fourier transforms to invert a 2-dimensional pgf for the joint distribution of cascade size and cumulative depth, allowing us to find the EATD. The inverse fast Fourier transform method is crucial for finding accurate distributions under the MTBP theory, other studies use computer algebra systems \cite{brummit2012}, which are limited by the number of terms that can be computed in the probability distribution and by the number of generations that can be accounted for in the branching process. The inability for computer algebra systems to recover cascade-size probabilities for branching processes that are not truncated after a short number of generations means that they can not accurately model many real-world processes that might survive beyond a small number of generations.
\par
In \cref{sec:app_to_data} of this paper, we applied the MTBP method to synthetic and real-world networks. Using the edge-disjoint edge clique cover (EECC) method from Burgio et al.~\cite{burgio21}, we recovered the Newman-Miller distribution of both a synthetically generated network (\cref{fig:EECC_comparison_synthetic}) and a Newman-Miller approximation for the clique distribution of three real-world networks (\cref{fig:real_world_pdfs} and \cref{fig:real_world_ccdfs}) and compared the cascade-size distribution from the MTBP theory for given complex contagion dynamics to a simulation of the dynamics on the empirical network. For the synthetically generated network, the match between the theory and simulations is very close with only small finite-size effects --- in MTBPs we assume that the network is infinitely large. The MTBP theory better models the dynamics for complex contagion, but for simple contagion dynamics, there is very little difference in the accuracy of the theory using simple branching processes in comparison to MTBPs, the simple branching process is very accurate despite the network violating the locally tree-like assumption. For the real-world networks, we also get a good match between the theory and simulations for small cascade sizes (cascades up to size 3 account for over $99\%$ of cascades); however, for larger cascade sizes the MTBP is less accurate, this is likely because the clique cover decomposes larger cliques into a combination of triangles and single links. If very large cliques are present in the network when there is social reinforcement present; i.e., $\alpha>0$, the simulated cascades can be much larger than those predicted by the theory. In all real-world networks that we show here, the MTBP is more accurate in predicting the cascade size distribution than a simple branching process approximation; however, an issue with using real-world networks is that if they are too small we often encounter finite-size effects (which we explore further in \cref{sec:finite_size}) and if they are too large the EECC algorithm can take too long to converge.
\par
The examples here were restricted to Newman-Miller networks, where we assume that the cliques never have more than three nodes; however, in many real-world networks, this is an over-simplification and not accounting for larger cliques can lead to less accurate results. If future models were to account for these larger cliques, we would expect more accurate results. We have not extended the model beyond cliques of size three as deriving the pgfs for quantities such as cascade size by hand to include very large clique sizes would take a very long time. We imagine that it would be possible to write a computer algorithm to derive the equations; however, this is beyond the scope of this paper. We also assumed that the cliques do not share edges; i.e., are edge-disjoint, in reality this is not always the case; however, it is possible to extend the model to account for this by including motif types which are arrangements of three cliques which share edges.

\appendix

\section{Simple branching process approximation for the cascade size calculation\label{sec:SED_cascade_size}}

For comparison with the results that account for triangles, here we show how to approximate the dynamics with a single-edge decomposition (SED); i.e., assuming that the network is locally tree-like. As in \cref{sec:size_dist}, let $\tilde{X}_{n}$ be the size of a full cascade allowed to grow for $n$ generations; then, we can write $\tilde{X}_{n}$ in terms of the sizes of its subtrees,
\begin{equation}
    \tilde{X}_{n}=1+\sum_{i}X_{n}^{(i)},
\end{equation}
where $X_{n}^{(i)}$ is the size of subtree $i$ allowed to grow for $n$ generations. Similarly to in \cref{sec:size_dist}, the size of a subtree does not include its seed node, in the SED there is only one subtree type and it is the same as a type-5 subtree in \cref{sec:size_dist}. The pgf for cascade size is
\begin{align}
    \tilde{K}_{n}(z)&=\mathbb{E}\left[z^{\tilde{X}_{n}}\right]\label{eq:sed1}\\
    &=\sum_{m}\pi_{m}\mathbb{E}\left[z^{1+\sum_{i=1}^{m}X_{n}^{(i)}}\right]\nonumber\\
    &=\sum_{m}\pi_{m}z\mathbb{E}\left[z^{X_{n}}\right]^{m}\nonumber\\
    &=z\tilde{f}\left(K_{n}(z)\right)\nonumber,
\end{align}
where $\pi_{m}$ is the probability that a node chosen at random has degree $m$ and $\tilde{f}(z)$ is the pgf for the degree distribution. Similarly,
\begin{align}
    K_{n}(z) &= \mathbb{E}\left[z^{X_{n}}\right]\label{eq:sed2}\\
    &=1-p_{1}+p_{1}\mathbb{E}\left[z^{1+\sum_{i}X_{n-1}^{(i)}}\right]\nonumber\\
    &=1-p_{1}+p_{1}z\sum_{m}q_{m}\mathbb{E}\left[z^{X_{n-1}}\right]^{m}\nonumber\\
    &=1-p_{1}+p_{1}zf\left(K_{n-1}(z)\right)\nonumber,
\end{align}
where $q_{m}$ is the probability that the excess degree of a node reached by traversing a link is $m$ and $f(z)$ is the pgf for the excess degree distribution. To find the full pgf at the point $z$, we iterate through equations \cref{eq:sed1} and \cref{eq:sed2} with initial conditions $K_{0}(z)=1$ until $K_{n}(z)$ is within a tolerance of $10^{-5}$ of $K_{n-1}(z)$ when evaluated for a specific value for $z$. We then find the full probability distribution from the pgf using the method described in \cref{section:1D_inversion}.

\section{EECC algorithm with maximum clique size 3\label{sec:EECC_algorithm}}

In this section, we detail the EECC algorithm that we use in \cref{sec:app_to_data}, this algorithm is a modification of the EECC introduced by \cite{burgio21} which was suggested but not explicitly presented in the paper \cite{burgio21}. In our modification, we restrict the clique size to a maximum of 3, naturally this can be extended to larger maximum sizes. The steps are as follows:
\begin{enumerate}
    \item Let $G$ be the network, find $C$, the set of maximal cliques in $G$.
    \item Find all of the elements of C which have order not greater than 3 and do not share edges with other elements of $C$ and add these elements (cliques) to the EECC.
    \item Remove any elements from $C$ that are in the EECC.
    \item Let $C_{OL}$ be the set of all cliques in $C$ that share at least one edge with another clique in $C$ and let $C_{HO}$ be the set of all elements of $C$ that are of order greater than 3.
    \item Remove any elements from $C_{OL}$ that are also in $C_{HO}$.
    \item While $C_{OL}$ is not empty, for each element calculate $\rho$, the proportion of edges shared with other elements of $C_{OL}$. Add a randomly chosen element with the lowest $\rho$ to the EECC and remove from $C_{OL}$. Remove all edges present in the EECC from $C_{OL}$. Recalculate the maximal cliques and $\rho$ considering all remaining elements.
    \item While $C_{HO}$ is not empty, for each element calculate $\rho$, the proportion of edges shared with other elements of $C_{HO}$. For a randomly chosen element with the smallest value for $\rho$, if it is of order not greater than 3 add to the EECC; otherwise, add a subclique of size 3 to the EECC. Remove all edges present in the EECC from $C_{HO}$, recalculate the maximal cliques and $\rho$ considering all remaining elements.
    
\end{enumerate}

\section{Finite-size effects\label{sec:finite_size}}

In \cref{fig:real_world_pdfs} and \cref{fig:real_world_ccdfs}, we see differences between the cascade-size distribution for the Monte-Carlo simulations on the empirical networks and the cascade-size distributions from the MTBP theory, especially in the tails of the distributions. One potential cause for this is that, since the networks are not very large (the science co-authorship network has 379 nodes), these could be finite-size effects. When we model cascades on networks using branching processes, we make the assumption that the network is infinitely large; however, this assumption is clearly violated in the case of these real-world networks. Another possible cause of this differences between the theory and the simulations is that the real-world networks contain higher-order cliques and highly connected structures that are lost in the MTBP model when we approximate the network structure using the EECC with maximum clique size of three.
\par
\begin{figure}[h!]
    \centering
    \begin{subfigure}{0.45\textwidth}
         \centering
         \includegraphics[width=\textwidth]{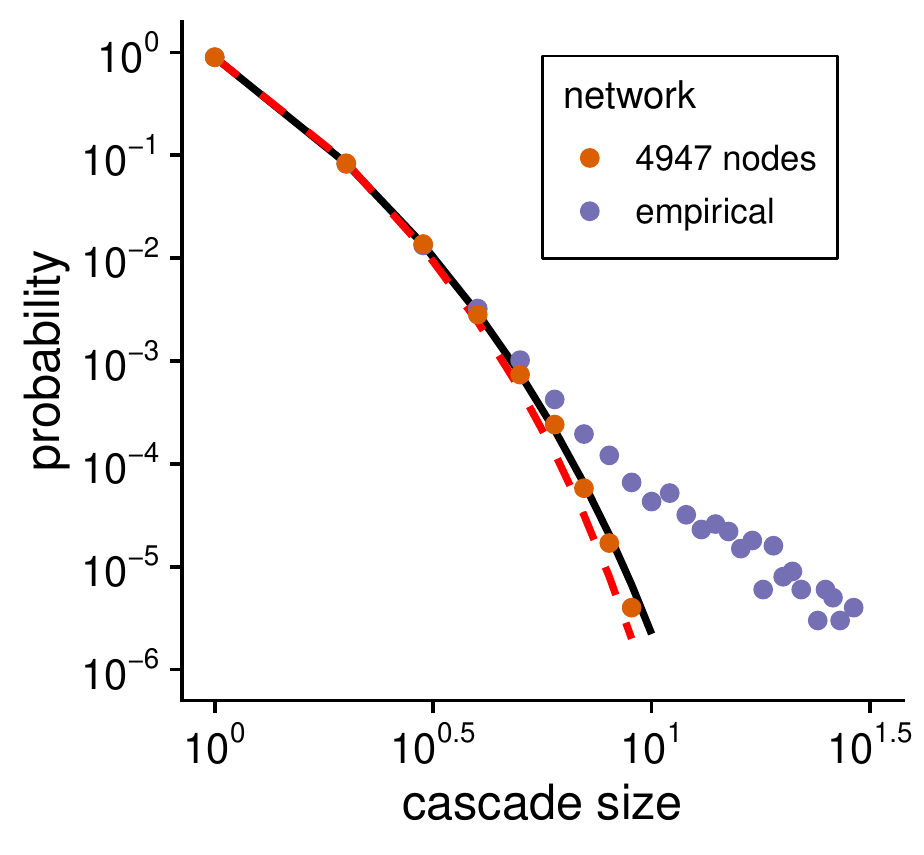}
         \caption{powergrid: probability distribution}
     \end{subfigure}
    \begin{subfigure}{0.45\textwidth}
         \centering
         \includegraphics[width=\textwidth]{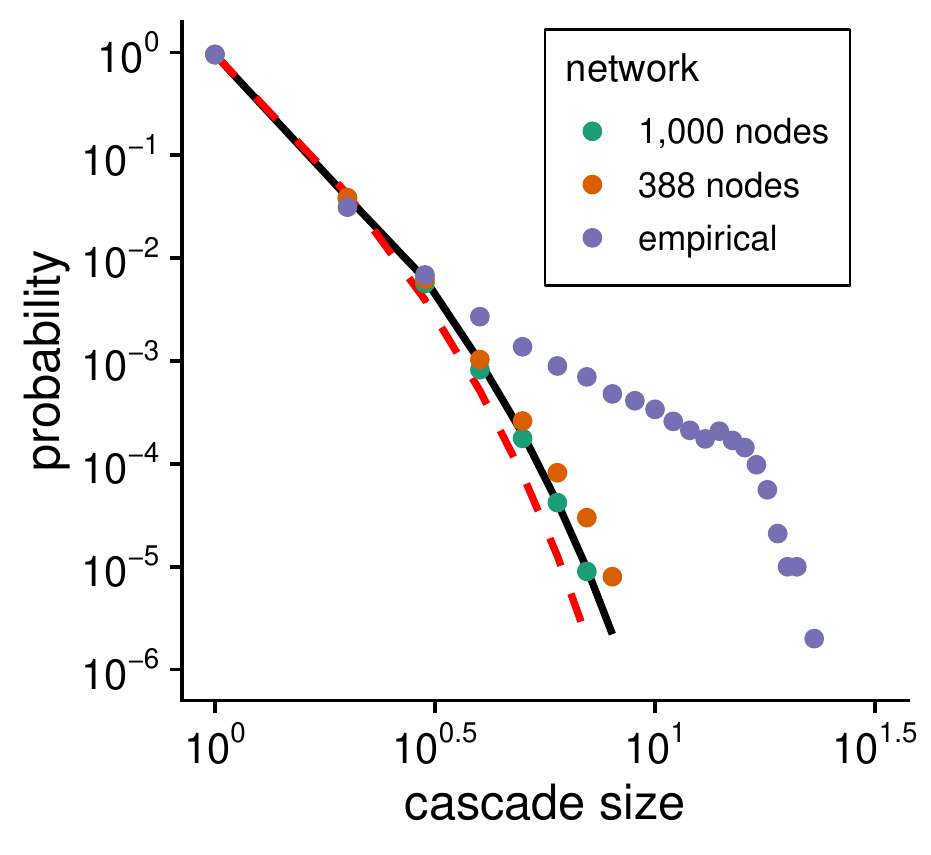}
         \caption{science co-authorship: probability distribution}
     \end{subfigure}
    \begin{subfigure}{0.45\textwidth}
         \centering
         \includegraphics[width=\textwidth]{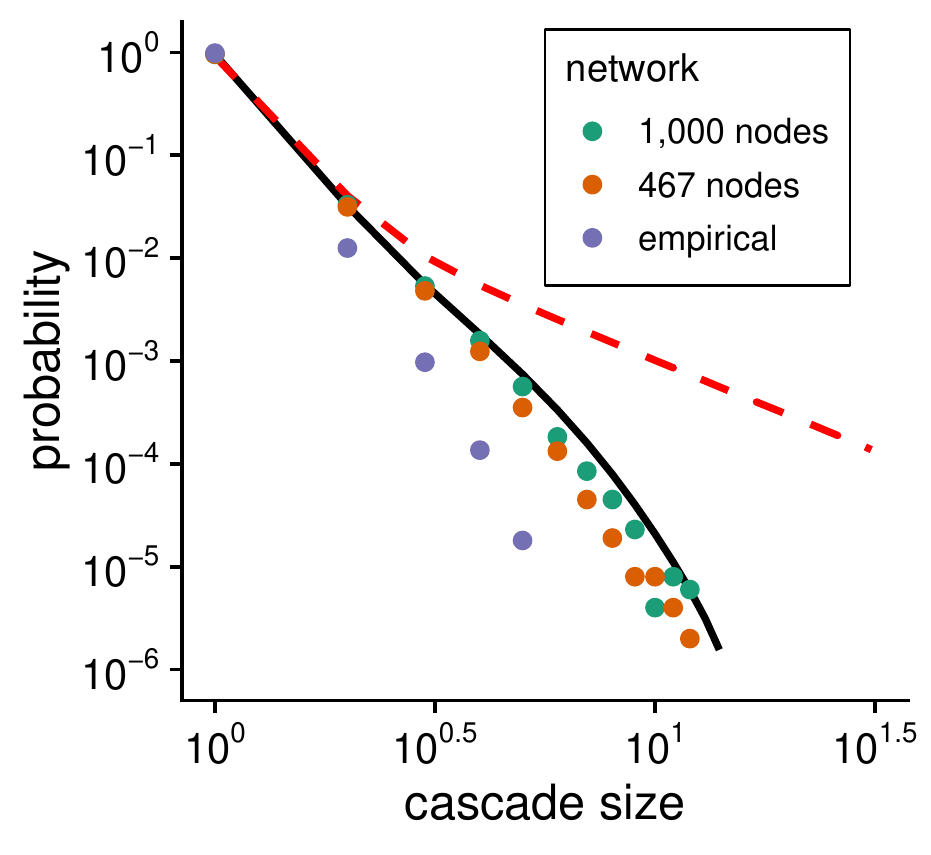}
         \caption{C.~elegans: probability distribution}
     \end{subfigure}
    \caption{The aim of this figure is to explore the impact of finite-size effects in the Monte-Carlo simulations. For each of the real-world networks studied, we constructed a synthetic Newman-Miller network of approximately the same size as the empirical network using the Newman-Miller distribution retrieved from finding an EECC on the empirical network. We performed 1 million Monte-Carlo simulations on each of these networks. For each of these networks, the cascade-size distribution on the empirical network is shown by the purple points and the simulations on the synthetic network of approximately the same size is shown by the orange points. The closeness between the orange points and the theory suggests that the impact of finite-size effects is minimal in causing the difference between the theory and simulations on the empirical network.\label{fig:finite_size}}
\end{figure}

All cliques with more than three nodes are decomposed into a series of edge-disjoint triangles and single links. To determine whether these discrepancies are predominantly due to finite-size effects, we generate a synthetic network of approximately the same size as the empirical networks from the Newman-Miller distributions calculated from their EECC and run 1 million Monte-Carlo simulations on each of these synthetic networks. These synthetic networks meet the assumption that is made in the theory that the network is fully composed of triangles and single links. Therefore, we can disentangle the impact of finite-size effects from other structural features present in the empirical network which are not within the assumptions of the MTBP theory. In \cref{fig:finite_size}, we show the cascade-size distributions for the powergrid network \cite{watts1998collective}, the largest connected component of the science co-authorship network \cite{Newman2006} and the C.~elegans network \cite{duch2005} for both the MTBP theory (black solid line), the simulations on the empirical network (purple points) and simulations on a synthetic network for a similar size generated from the Newman-Miller distribution obtained from the EECC on that network (orange points).
\par
We find that the differences between the theory and the simulations on the synthetic network are minimal in comparison to the differences between the theory and the simulations on empirical networks. This confirms that the impact of finite-size effects is minimal for the simulations and that the discrepancies between the distributions from the MTBP theory and the Monte-Carlo simulations on real-world networks are most likely due to the fact that there are differences other than size in the network structure between the empirical network and in the distribution that we use in the MTBP from the EECC.

\FloatBarrier
\bibliographystyle{siamplain}
\bibliography{references}

\end{document}


\maketitle

\section{Simple branching process approximation for the cascade size calculation\label{sec:SED_cascade_size}}

We follow the method used in Ref.~\cite{gleeson2021} to find the cascade size distribution under a simple branching process approximation; i.e., ignoring the clustering. We are interested in calculating the distribution under a simple branching process approximation for comparison with the MTBP, both cascade size distributions are shown in \cref{fig:cascade_size_dist}.\par
Gleeson et al.~\cite{gleeson2021} derived the following equations for finding the cascade size pgf for a cascade allowed to grow for $n$ generations $\Tilde{G}_{n}(x)$;
\begin{equation}
    \Tilde{G}_{n}(x)=x\Tilde{f}(G_{n-1}(x))\label{eq:simplebp1}
\end{equation}
and
\begin{equation}
    G_{n}(x) = xf(G_{n-1}(x))\label{eq:simplebp2}
\end{equation}
where $\Tilde{f}(x)$ and $f(x)$ are the pgfs for the number of offspring from a seed node and a node that is not a seed respectively. \Cref{eq:simplebp1,eq:simplebp2} can be solved by iterating with initial condition $G_{0}(x)=x$. In Ref.~\cite{gleeson2021} they show how to find $\Tilde{f}(x)$ and $f(x)$ from data, here we show how to find these quantities from the pgf for the network degree distribution.\par
Let $\Tilde{Z}$ be the number of offspring from the the seed node, then its pgf $\Tilde{f}(x)$ is given by

\begin{align}
    \Tilde{f}(x) &= \mathbb{E}\left[x^{\Tilde{Z}}\right]\\
    &=\sum_{k=0}^{\infty}\pi_{k}\mathbb{E}\left[x^{\Tilde{Z}}\middle| \text{ node is degree }k\right]\nonumber\\
    &=\sum_{k=0}^{\infty}\pi_{k}\sum_{l=0}^{k}\binom{k}{l}p_{1}^{l}(1-p_{1})^{k-l}x^{l}\nonumber\\
    &=\sum_{k=0}^{\infty}\pi_{k}\sum_{l=0}^{k}\binom{k}{l}(p_{1}x)^{l}(1-p_{1})^{k-l}\nonumber\\
    &=\sum_{k=0}^{\infty}\pi_{k}(p_{1}x + 1 - p_{1})^{k}\nonumber\\
    &=\Tilde{h}(p_{1}x + 1 - p_{1})\nonumber
\end{align}

where $\Tilde{h}(x)$ is the pgf for the degree distribution of the network and $p_{1}$ is the adoption probability. Similarly, it can be shown that
\begin{equation}
    f(x) = h(p_{1}x + 1 - p_{1}).
\end{equation}
In this paper, our focus is on distributions of triangles and single links, Newman \cite{Newman2009} showed that
\begin{equation}
    \Tilde{h}(x) = \tilde{P}(z,z^{2})
\end{equation}
where $P(x,y)$ is the joint distribution of triangles and single links.

\bibliographystyle{siamplain}
\bibliography{references}